\documentclass[aps,prd,reprint,superscriptaddress,nofootinbib]{revtex4-2}

\usepackage{amsmath,amssymb,graphicx,bm,physics,hyperref}
\usepackage{ulem}
\usepackage[utf8]{inputenc}
\usepackage[english]{babel}
\usepackage{xcolor}
\makeatletter
\expandafter\let\csname l@en\endcsname\l@english
\makeatother

\begin{document}

\title{Diffraction by Circular and Triangular Apertures\\ as a Diagnostic Tool of Twisted Matter Waves}

\author{M.~Maksimov}
\email{maksim.maksimov@metalab.ifmo.ru}
\affiliation{School of Physics and Engineering, ITMO University, 197101 St. Petersburg, Russia}

\author{N.~Borodin}
\affiliation{Joint Institute for Nuclear Research, Dubna, Russia}

\author{D.~Kargina}
\affiliation{School of Physics and Engineering, ITMO University, 197101 St. Petersburg, Russia}
\affiliation{Petersburg Nuclear Physics Institute of NRC ``Kurchatov Institute'', Gatchina, Russia}

\author{D.~Naumov}
\affiliation{Joint Institute for Nuclear Research, Dubna, Russia}

\author{D.~Karlovets}
\affiliation{School of Physics and Engineering, ITMO University, 197101 St. Petersburg, Russia}
\affiliation{Petersburg Nuclear Physics Institute of NRC ``Kurchatov Institute'', Gatchina, Russia}

\begin{abstract}
We study diffraction of twisted matter waves—electrons and light ions carrying orbital angular momentum $\ell/\hbar = 0, \pm 1, \pm 2, \ldots$—by circular and triangular apertures. Within the scalar Kirchhoff–Fresnel framework, circular apertures preserve cylindrical symmetry and produce ringlike far-field profiles whose radii and widths depend on $|\ell|$ but are insensitive to its sign. In contrast, equilateral triangles break axial symmetry and yield structured patterns that encode both the magnitude and the sign of $\ell$. A transparent Fraunhofer mapping links detector coordinates to the Fourier plane, explaining the $(|\ell|+1)$-lobe rule and the sign-dependent rotation of the pattern. We validate these results for both ideal Bessel beams and localized Laguerre–Gaussian packets, and we cross-check them by split-step Fourier propagation of the time-dependent Schr\"odinger equation. From these analyses we extract practical design rules—Fraunhofer distance, lattice pitch, and detector sampling—relevant to OAM diagnostics with moderately relativistic electrons of $E_{\text{kin}}\!\sim\!0.1$–$5$ MeV and light ions of $E_{\text{kin}}\!\sim\!0.1$–$1$ MeV/u. Our results establish triangular diffraction as a simple, passive, and robust method for reading out the OAM content of structured quantum beams.
\end{abstract}

\maketitle

\section{Introduction}\label{sec:intro}
Quantum states carrying orbital angular momentum (OAM)—often called \textit{twisted} states—have attracted sustained interest across optics~\cite{Krenn_2014,willner2022OAM}, atomic and molecular physics~\cite{Session2025,luski2021vortex}, and electron microscopy~\cite{allen1999orbital,bliokh2017vortex,larocque2018twisted,FloettmannKarlovets2020}. These modes, with helical phase $\exp(i\ell\varphi)$ and quantized projection $\ell\hbar$, enable sensitive probes of interference and scattering; twisted photons from transition radiation have also been observed~\cite{Takabayashi2025TRAOM}. In electron optics, vortex beams have been generated and controlled in TEM~\cite{uchida2010nature,verbeeck2010nature,guzzinati2014,grillo2016bessel}, opening avenues for OAM-resolved imaging and spectroscopy~\cite{karlovets2021vortex}. Related efforts extend to twisted atoms and ions~\cite{luski2021vortex}. Beyond microscopy, OAM is relevant in neutron optics~\cite{Sarenac2022SciAdv}, accelerator physics~\cite{FloettmannKarlovets2020,karlovets2021vortex}, and high-energy physics~\cite{bliokh2017vortex,Ivanov2022PPNP}.

Aperture diffraction provides a simple, noninterferometric diagnostic of a beam’s transverse phase~\cite{goodman2005fourier}. Circular apertures preserve axial symmetry and are therefore insensitive to the sign of $\ell$, with only a weak $|\ell|$-dependence via ring radii and widths. In contrast, equilateral triangular apertures break rotational symmetry and generate far-field patterns that encode both the magnitude and the sign of $\ell$~\cite{hickmann2010unveiling,stahl2019analytic,ambuj2019symmetry,juchtmans2016friedel}. Variants include annular masks, line apertures, and spiral-imaging schemes~\cite{annularTriangle2021,hasan2025hashtype,jiang2015spiralimaging}.

Here we analyze diffraction of twisted matter waves—electrons and light ions—by circular and triangular apertures within the scalar Kirchhoff–Fresnel framework, and we validate the analysis by split-step Fourier propagation of the time-dependent Schr\"odinger equation. We consider both ideal Bessel inputs and localized Laguerre–Gaussian (LG) packets over nonrelativistic to moderately relativistic energies. Emphasis is placed on practical scalings that turn triangular diffraction into a passive OAM diagnostic. At ultrarelativistic energies, a key limitation is the shrinking lattice pitch in the far field, which tightens detector-resolution requirements and lengthens the propagation distances needed to reach a clean Fraunhofer regime. The geometries we study are directly relevant to the ongoing experiments within the ITMO–JINR program~\cite{ivanov2023vortex,PhysRevA.110.L031101,Dyatlov2025DUV,RSF2023} and to the future work on twisted ions at the Institute of Modern Physics in China~\cite{An_2025}.

Our main results are: (i) a unified, symmetry-based explanation of the triangular far-field pattern with explicit dependencies on the sign and magnitude of $\ell$; (ii) compact design rules and scalings with wavelength, aperture size, and propagation distance for forming a clear far-field lattice; (iii) a circular-aperture benchmark that confirms insensitivity to $\mathrm{sign}(\ell)$ while quantifying the residual $|\ell|$-dependence; and (iv) validation of robustness for realistic LG packets—including dispersive ion beams—by cross-checking analytics against direct Kirchhoff evaluation and split-step simulations.

The paper is organized as follows. Section~\ref{sec:theory} introduces the Kirchhoff–Fresnel formalism and the Fraunhofer mapping, defines Bessel and LG incident states, and analyzes the triangular mask. Section~\ref{sec:numerics} details the paraxial FFT propagator, aperture modeling, and count-rate normalization. Section~\ref{sec:results} present (i) a circular-aperture benchmark, (ii) triangular diffraction for electrons with $z$-scans and energy scaling up to moderately relativistic, (iii) Bessel vs.\ LG comparisons, and (iv) ion cases. Section~\ref{sec:feasibility} collects practical criteria (far-field condition, lattice pitch, detector resolution, and event-rate estimates). Sections ~\ref{sec:discussion}-~\ref{sec:conclusion} conclude our work with limitations and outlook.

\section{Theory and Methods}\label{sec:theory}

\subsection{Schr\"odinger equation and Kirchhoff--Fresnel integral}\label{subsec:sch-kf}
We work in Cartesian coordinates $(x,y,z)$ with the optical axis along $z$; cylindrical $(\rho,\phi,z)$ appear only when symmetry warrants it. Free evolution of quantum particles such as electrons or ions carrying OAM is generally described by the Klein–Gordon equation. In the nonrelativistic limit this reduces to the familiar time-dependent Schrödinger equation
\begin{equation}
i\hbar\, \frac{\partial}{\partial t} \, \psi(\mathbf{r}, t) = -\frac{\hbar^2}{2m} \nabla^2 \psi(\mathbf{r}, t),
\label{eq:tdse_1}
\end{equation}
with $E_{kin} = p^2/2m$. When relativistic kinematics matters, we retain the exact dispersion law $E = \sqrt{p^2 c^2 + m^2 c^4}$.

\begin{figure*}[t]
  \centering
  \includegraphics[width=1\textwidth]{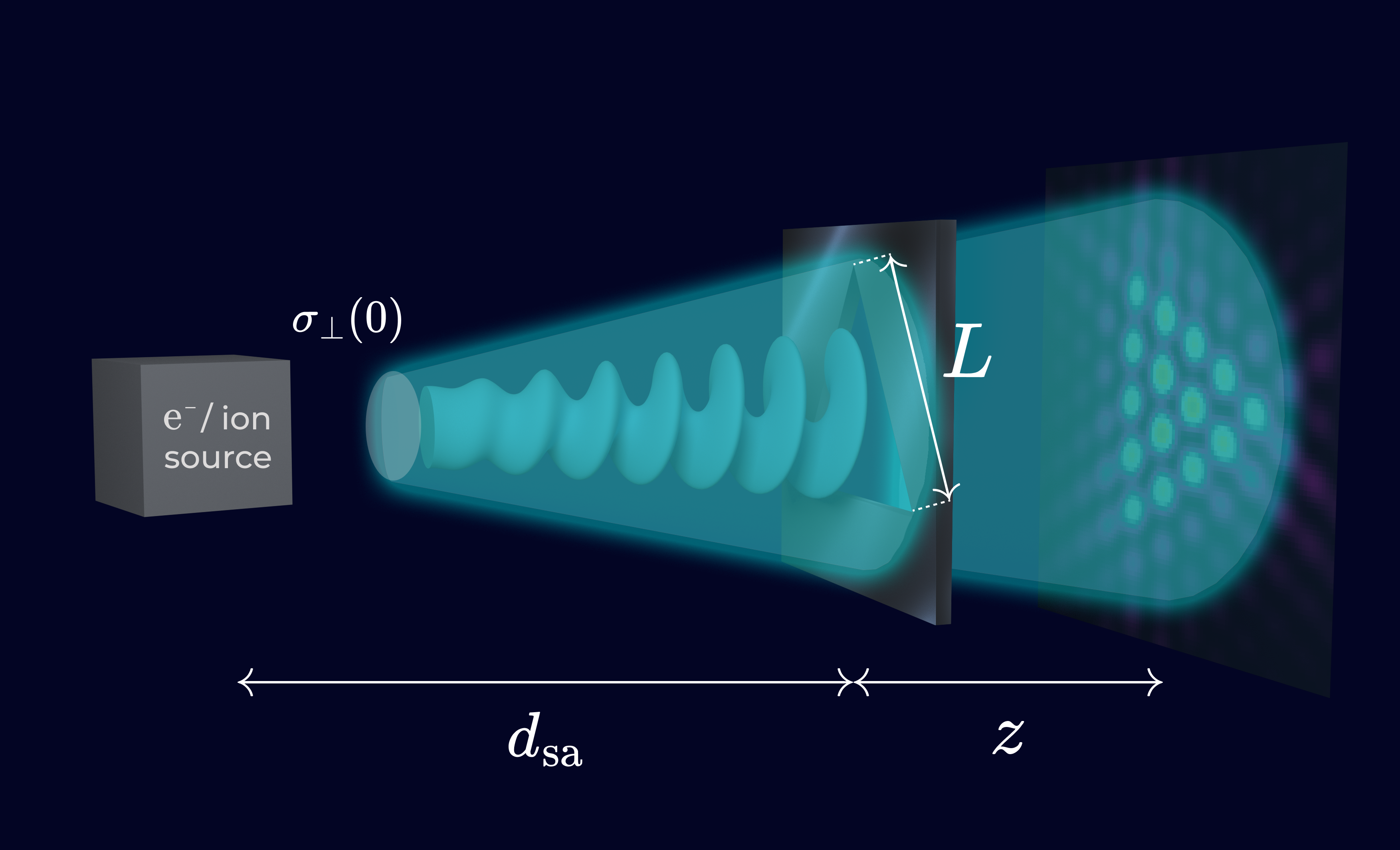}
  \caption{Schematic of the diffraction geometry for twisted matter waves: source $\rightarrow$ equilateral triangular aperture (side $L$) at $z=0$ $\rightarrow$ detection screen at distance $z$. The source–aperture distance is $d_{\mathrm{sa}}$, and $\sigma_{\perp}(0)$ denotes the initial transverse beam size. Not to scale.}
  \label{fig:setup_triangle}
\end{figure*}

We consider diffraction of twisted wave packets by a perfectly absorbing screen in the plane $z=0$ with an open aperture area S of arbitrary shape. The same formalism covers a circle of radius $a$ and an equilateral triangular opening of side $L$. See Fig.~\ref{fig:setup_triangle} for a schematic of the geometry with triangular aperture.

For a monochromatic component with wavenumber $\mathbf k=|\mathbf p|/\hbar$, the spatial field obeys the scalar Helmholtz equation
\begin{equation}
\left(\nabla^2+k^2\right)\psi(\mathbf r)=0.
\end{equation}
Applying Green’s second identity under the standard Kirchhoff boundary conditions yields the scalar Kirchhoff–Fresnel diffraction integral~\cite{jackson1999classical,goodman2005fourier}
\begin{equation}
\psi(\mathbf r)=
\int_{S}\frac{e^{ikR}}{4\pi R}\,
\mathbf n'\cdot\bigg[
\mathbf \nabla'\psi(\mathbf r')+\Big(ik-\frac{1}{R}\Big)\frac{\mathbf R}{R}\,\psi(\mathbf r')
\bigg]\; dS',
\label{eq:kirchhoff}
\end{equation}
where the observation point is $\mathbf r=(x,y,z)$, the integration point is $\mathbf r'=(x',y',z')$ on a generic surface $S$ with unit normal $\mathbf n'$, $\mathbf R=\mathbf r-\mathbf r'$, and $R=|\mathbf R|$. Choosing $S$ as a circle or as a triangle recovers the circular and triangular masks used below.

Although historically derived in classical optics, Eq.~\eqref{eq:kirchhoff} applies to massive quantum particles provided the description is scalar, propagation occurs in free space without external fields, and the absorbing screen is well described by the Kirchhoff boundary conditions, with spin remaining uncoupled and the monochromatic component selected from the packet’s Fourier decomposition.
In this form, the integral can be employed in diffraction studies of electrons, muons, neutrons, ions, and so forth.

\subsection{Fraunhofer mapping and validity}\label{subsec:fraunhofer-mapping}
When a vortex beam passes through a triangular aperture, the far-field (Fraunhofer) diffraction pattern develops a striking array of bright spots. A natural question arises: what governs the number, the positions, and the overall arrangement of these spots? In what follows we derive the mapping from the aperture field to the far zone, analyze the Fourier transform of the triangular mask, and explain how the vortex phase selects a finite set of reciprocal-lattice nodes that manifest as the observed spot pattern.

In the far-field zone the Kirchhoff–Fresnel integral over the planar screen $S$ at $z'=0$ can then be simplified to~\cite{jackson1999classical}
\begin{equation}
\label{eq:KF}
\psi(x,y,z)\;\propto\; \iint_{S} \psi(x',y')\,
\frac{e^{i k R}}{R}\;dx'\,dy',
\end{equation}
where \(R=\sqrt{(x-x')^2+(y-y')^2+z^2}\)
For a characteristic aperture size $D$, the Fraunhofer regime holds when
\begin{equation}
\label{eq:fraunhofer}
z\;\gtrsim\;z_{F} = \frac{D^2}{\lambda_{\mathrm{dB}}}, \qquad \lambda_{\mathrm{dB}} = \frac{2\pi\hbar}{\langle p \rangle}.
\end{equation}

One can expand $R$ for large $z$ and, after neglecting the weak quadratic phase across a small aperture, obtain
\begin{equation}
\label{eq:Fraunhofer-mapping}
\begin{aligned}
\psi(x,y,z) \;& \propto\;
\frac{e^{ik\left(z+\tfrac{x^2+y^2}{2z}\right)}}{z}\,
\iint_{-\infty}^{+\infty} dx'\,dy'\; e^{-i\frac{k}{z}\left(xx'+yy'\right)} \times\\
&\times A(x',y')\,\psi(x',y'),
\end{aligned}
\end{equation}
where $A(x',y')$ is the aperture transmission function (unity on the open set $S$, zero outside).
Therefore, the integral is the two-dimensional Fourier transform of $A(x',y')\,\psi(x',y')$ evaluated at
\begin{equation}
\label{eq:kxky-mapping}
k_x = \frac{k}{z}\,x = \frac{2\pi}{\lambda_{\mathrm{dB}}\, z}\,x,
\qquad
k_y = \frac{k}{z}\,y = \frac{2\pi}{\lambda_{\mathrm{dB}}\, z}\,y .
\end{equation}
The detector coordinates $(x,y)$ are linearly mapped to the Fourier plane $(k_x,k_y)$:
\begin{equation}
\label{eq:XY-from-k}
(x,y) = \frac{\lambda_{\mathrm{dB}}\, z}{2\pi}\,(k_x,k_y),
\end{equation}
so increasing $z$ magnifies the entire pattern without changing its shape; energy via $\lambda_{\mathrm{dB}}$ rescales the pattern uniformly. This mapping is the backbone of all scalings used below.

\subsection{Incident twisted beams: Bessel vs.\ Laguerre--Gaussian}\label{subsec:incident-beams}
We model the incident OAM states with two complementary families: Bessel beams and localized LG packets.
\paragraph*{Bessel beams.}
A monochromatic Bessel beam with an azimuthal quantum number $\ell$ and longitudinal momentum $k_z$ has the form
\begin{equation}
\psi_\ell^{\text{B}}(\rho, \phi, z) = J_{|\ell|}\!\left( {\kappa \rho} \right) e^{i \ell \phi} e^{i k_z z}, \quad \kappa = \frac{p_\perp}{\hbar},
\label{eq:bessel_beam}
\end{equation}
where $J_{|\ell|}$ is the Bessel function of the first kind and $\kappa$ is the transverse momentum. These states carry $\langle L_z \rangle = \ell \hbar$ and exactly solve the free-space wave equation with fixed energy and momenta $(\kappa, k_z)$. Because $J_{|\ell|}(x) \sim x^{-1/2}$ at large $x$, the transverse norm diverges and such states are not physically normalizable. Nonetheless, they accurately capture the local structure of twisted wave fronts near the axis and are widely used as analytic inputs in diffraction theory~\cite{grillo2016bessel,bliokh2017vortex}.

\paragraph*{Laguerre--Gaussian wave packets.}
To describe physical, spatially localized OAM states, we use LG wave packets, which are exact square-integrable solutions of the Klein--Gordon equation for free particles with a well-defined mean energy and angular momentum~\cite{allen1999orbital,karlovets2021vortex}. Assuming a factorized form $\psi(\mathbf{r}, t) = \psi_\perp(\rho, \phi, t)\, \psi_z(z, t)$, the transverse part with a radial index $n=0,1,2,\dots$ and azimuthal index $\ell$ is
\begin{align}
\psi_{\ell,n}^{\text{LG}}(\rho, \phi, t) &= \mathcal{N} \, \rho^{|\ell|} \left[ \frac{1}{\sigma_\perp(t)} \right]^{|\ell|+1}
L_n^{|\ell|} \left( \frac{\rho^2}{\sigma_\perp^2(t)} \right) e^{i \ell \phi} \notag \\
&\quad \times \exp\left[ -\frac{\rho^2}{2\sigma_\perp^2(t)}\left(1 - i \frac{t}{t_d} \right) \right] \notag \\
&\quad \times \exp\left[ -i M \arctan\left( \frac{t}{t_d} \right) \right],
\label{eq:LG_packet}
\end{align}
where $L_n^{|\ell|}$ is the associated Laguerre polynomial and $\mathcal{N}$ is a normalization constant and
\begin{equation}
M = 2n + |\ell| + 1,
\label{eq:M_def}
\end{equation}
which in the accelerator and beam-physics literature plays the role of a beam-quality factor ~\cite{karlovets2021vortex}.

The initial transverse width $\sigma_\perp(0)$ is related to the rms transverse momentum spread $\sigma_p$ as
\begin{equation}
\sigma_\perp(0) = \frac{1}{\sigma_p}, \quad
t_d = \frac{m}{\sigma_p^2}.
\label{eq:LG_defs}
\end{equation}

The width evolves as
\begin{equation}
\sigma_\perp(t) = \sigma_\perp(0) \sqrt{1 + \left( \frac{t}{t_d} \right)^2 } = \sigma_\perp(0) \sqrt{1 + \left( \frac{\langle z \rangle}{z_R} \right)^2 },
\label{eq:LG_width}
\end{equation}
with the Rayleigh length, mean velocity
\begin{equation}
z_R = \frac{2\pi}{M} \frac{\sigma_\perp^2(0)}{\lambda_{\mathrm{dB}}},
\quad
\langle u \rangle = \frac{\langle z \rangle}{t}
\label{eq:zR}
\end{equation}
and a Gouy phase
\begin{equation}
\Phi_{\text{Gouy}}(t) = M \arctan\left( \frac{t}{t_d} \right).
\end{equation}
The normalization
\begin{equation}
\int_0^\infty \rho\, d\rho \int_0^{2\pi} d\phi\, |\psi_{\ell,n}(\rho, \phi, t)|^2 = 1
\end{equation}
ensures $\langle L_z \rangle = \ell \hbar$ for any t. For $\ell \ne 0$ the intensity has a central minimum and a ring-like vortex profile.

\subsection{Spreading scales}

For electron beams with initial width $\sigma_\perp(0)\sim 1$~nm and kinetic energies of $100$--$5000$~keV, Eq.~\eqref{eq:zR} yields typical Rayleigh lengths of $z_R\sim 1.5$--$30~\mu\mathrm{m}$~\cite{karlovets2021vortex}.  
For heavier particles at the same $\sigma_p$, the smaller de~Broglie wavelength $\lambda_{\mathrm{dB}}$ would increase $z_R$ if all other parameters remained fixed. In practice, however, larger mass usually correlates with smaller attainable $\sigma_\perp(0)$, and since $z_R \propto \sigma_\perp^2(0)$, this dominates and leads to shorter Rayleigh lengths.  

In what follows we restrict to paraxial packets,
\begin{equation}
\sigma_p \ll p_z \quad \Leftrightarrow \quad \sigma_\perp(0) \gg \lambda_{\mathrm{dB}},
\end{equation}
so that the longitudinal momentum is well defined and diffraction patterns remain stable during free propagation.  

To quantify transverse coherence at an observation plane, we use the rms transverse radius squared
\begin{equation}
\langle \rho^2 \rangle (t) = \sigma_\perp^2(t)\,M.
\label{eq:xi_def}
\end{equation}
In the far field, $\langle z \rangle \gg z_R$, the rms radii at the source and observation plane are related by van Cittert-Zernike theorem~\cite{karlovets2021vortex}
\begin{equation}
\sqrt{\langle \rho^2 \rangle}(\langle z \rangle) \;=\;
\frac{\langle z \rangle\, \lambda_{\mathrm{dB}}}{2\pi \sqrt{\langle \rho^2 \rangle}(0)} \, M.
\label{eq:karlovets_relation}
\end{equation}
As a practical criterion for preserving OAM-dependent diffraction features after transmission through an aperture, one requires
\begin{equation}
\sqrt{\langle \rho^2 \rangle}(d_{\rm sa}) \;\gtrsim\; D ,
\label{eq:xi_vs_L}
\end{equation}
where $d_{\rm sa}$ is a distance between the source and the aperture and $D$ is the effective aperture size.

\subsection{Triangular mask: analysis and symmetry}\label{subsec:triangle-analysis}
A triangular aperture resolves both the magnitude and the sign of the OAM. In the far field, the Fraunhofer mapping relates detector and Fourier coordinates, so the structure of the spectrum $\widetilde{A}(k_x,k_y)$ of the mask controls the observed pattern, where $\widetilde{A}$ is the Fourier transform of $A(x',y')$ as defined in App.~\ref{app:triangle-derivations}, Eq.~\eqref{eq:TriFT-start_app}. For a vortex input $\psi_{\rm in}(x,y)\propto \rho^{|\ell|} e^{i\ell\phi}$, the incident helical phase acts as an $|\ell|$-th order differential operator in $\mathbf{k}$-space, producing a finite sum of transverse derivatives of the mask spectrum and thereby emphasizing its rapidly varying regions as bright spots. The detailed derivation is deferred to App.~\ref{app:triangle-derivations}.

\paragraph*{Analytic Fourier amplitude of a triangular mask.}
Let the filled triangle in the aperture plane have vertices $\mathbf{v}_0,\mathbf{v}_1,\mathbf{v}_2$ and edge vectors $\mathbf{e}_1=\mathbf{v}_1-\mathbf{v}_0$, $\mathbf{e}_2=\mathbf{v}_2-\mathbf{v}_0$. Denote $\mathbf{k}=(k_x,k_y)$ and introduce the linear forms
\begin{equation}
\alpha = \mathbf{k}\!\cdot\!\mathbf{e}_1,\qquad
\beta  = \mathbf{k}\!\cdot\!\mathbf{e}_2,\qquad
\alpha-\beta = \mathbf{k}\!\cdot\!(\mathbf{e}_1-\mathbf{e}_2)
\label{eq:alpha_beta}
\end{equation}
and the Jacobian $J=|\mathbf{e_1}\times \mathbf{e_2}|$ which equals twice the area of the triangle. The mask’s Fourier amplitude admits a closed rational form
\begin{equation}
\label{eq:triangle-rational}
\widetilde{A}(\mathbf{k}) \;=\; J\,e^{-i\mathbf{k}\cdot \mathbf{v}_0}\;
\frac{\alpha\,e^{-i\beta}-\beta\,e^{-i\alpha}-(\alpha-\beta)}{\alpha\,\beta\,(\alpha-\beta)}\,.
\end{equation}
The factors $\alpha$, $\beta$, and $\alpha-\beta$ single out three preferred directions in $\mathbf{k}$-space, each orthogonal to one edge; along them $\widetilde{A}(\mathbf{k})$ develops alternating ridge lines. Their superposition produces the characteristic sixfold (hexagonal) symmetry of the far-field intensity.

\paragraph*{Reciprocal lattice and far-field spacing.}
Define the reciprocal basis vectors $\mathbf{G}_1,\mathbf{G}_2$ by $\mathbf{e}_i\!\cdot\! \mathbf{G}_j = 2\pi\delta_{ij}$. For an equilateral triangle of side $L$ one has $|\mathbf{G}_1|=|\mathbf{G}_2|=4\pi/(\sqrt{3}L)$ and $\angle(\mathbf{G}_1,\mathbf{G}_2)=60^\circ$. The three constructive-phase families
\[
\alpha=2\pi m,\qquad \beta=2\pi n,\qquad \alpha-\beta=2\pi p,\qquad m,n,p\in\mathbb{Z},
\]
intersect at the reciprocal-lattice nodes
\begin{equation}
\label{eq:kmn}
\mathbf k_{m,n} = m\,\mathbf G_1 + n\,\mathbf G_2 , \qquad m,n\in\mathbb Z,
\end{equation}
which map to detector coordinates by the far-field relation. The nearest-neighbour spacing along a reciprocal-basis direction translates into the detector-plane step
\begin{equation}
\label{eq:Delta-step}
\Delta \;=\; \frac{\lambda_{\mathrm{dB}} z}{2\pi}\,|G_1| \;=\; \frac{2\,\lambda_{\mathrm{dB}}\, z}{\sqrt{3}\,L}\,.
\end{equation}
The $|\ell|$-th order differential selection emphasizes a finite set of such nodes; in particular, the nodes with nonnegative $(m,n)$ and $m+n\le |\ell|$ are highlighted, and the global orientation of the triangular motif flips under $\ell\to-\ell$.

\section{Numerical implementation}\label{sec:numerics}
To validate and cross-check the results obtained with the Kirchhoff integral, we employed the split--step Fourier method (SSFM) --- a simple, efficient, and widely used numerical technique~\cite{weideman1986split}.  

We start from the time-dependent Schr\"odinger equation with a potential $V(\mathbf r)$
\begin{equation}
i\hbar\,\frac{\partial}{\partial t}\psi(\mathbf r,t) \;=\;
\left[-\,\frac{\hbar^2}{2m}\,\nabla^2 + V(\mathbf r)\right]\psi(\mathbf r,t),
\label{eq:tdse_2}
\end{equation}
whose formal solution over a time step $\Delta t$ can be written as
\begin{equation}
\psi(t+\Delta t) = U(\Delta t)\,\psi(t),\;\;
U(\Delta t) = \exp\!\left(-\tfrac{i}{\hbar}\hat H \Delta t\right),
\label{eq:ssfm_U}
\end{equation}
where $\hat H=\hat T+\hat V$. Here $\hat T$ and $\hat V$ denote the kinetic and potential-energy operators, with
\begin{equation}
\hat T \;=\; -\,\frac{\hbar^2}{2m}\,\nabla^2,
\qquad
\big[\hat V\psi\big](\mathbf r)=V(\mathbf r)\,\psi(\mathbf r).
\label{eq:ssfm_T}
\end{equation}

The idea of SSFM is to approximate the exact time-evolution operator by the Strang splitting,
\begin{equation}
U(\Delta t)\;\approx\;
e^{-\,\tfrac{i}{2\hbar} \hat V \Delta t}\,
e^{-\,\tfrac{i}{\hbar} \hat T \Delta t}\,
e^{-\,\tfrac{i}{2\hbar} \hat V \Delta t},
\label{eq:ssfm_Strang}
\end{equation}
which is accurate to $O(\Delta t^3)$.

Each exponential factor is applied in the representation where it is diagonal:  
the potential term $\hat V$ in real space, and the kinetic term $\hat T$ in Fourier space.  
This results in an efficient and strictly unitary algorithm with computational cost scaling as $N \log N$~\cite{weideman1986split}.  
In principle, the SSFM can treat arbitrary potentials and multiple scattering events.  
In practice, however, for electrons at keV energies the de~Broglie wavelength is only a few picometers, which makes a full three-dimensional grid numerically demanding.

For this regime, the paraxial approximation provides a practical simplification.  
We write the wave function as
\begin{equation}
\psi(\boldsymbol{\rho},z,t)\;\approx\;e^{i(kz-\omega t)}\,u(\boldsymbol{\rho},z),
\label{eq:ssfm_factorization}
\end{equation}
where in the nonrelativistic limit $\omega = \frac{\hbar k^2}{2m}$.  
This factorization separates out the rapid temporal oscillation, leaving a slowly varying envelope $u$ that depends only on the transverse coordinates $\boldsymbol{\rho}$ and the propagation distance $z$.  
The envelope satisfies
\begin{equation}
i\,\partial_z u(\boldsymbol{\rho},z) \;=\; -\,\frac{1}{2k}\,\nabla_\perp^2 u(\boldsymbol{\rho},z),
\label{eq:paraxial_SE}
\end{equation}
which is mathematically equivalent to a two-dimensional Schr\"odinger equation with $z$ playing the role of “time.” 
Thus the explicit time dependence disappears: physical detectors are far too slow to resolve oscillations on the scale of $\omega^{-1}$ and instead measure the stationary intensity.  

The numerical implementation is straightforward.  
For a propagation step of length $\Delta z$ the algorithm proceeds as follows:

\begin{enumerate}
\item Perform a forward Fast Fourier Transform (FFT) of the field $u(x,y,z)$, obtaining its transverse spectrum $\tilde u(k_x,k_y,z)$;
\item Multiply by the free-space propagator
\begin{equation}
U_T(k_x,k_y) \;=\; \exp\!\left[-\,\tfrac{i\Delta z}{2k}\,\bigl(k_x^2+k_y^2\bigr)\right];
\label{eq:ssfm_UT}
\end{equation}
\item Apply the inverse FFT to recover the updated field $u(x,y,z+\Delta z)$.
\end{enumerate}

Within the paraxial framework, apertures and other thin elements are included as multiplicative masks $A(x,y)\in[0,1]$ at selected $z$-planes, playing the role of the potential:
\begin{equation}
u(x,y,z_0^+)\;=\;A(x,y)\,u(x,y,z_0^-).
\label{eq:ssfm_mask}
\end{equation}
An ideal sharp aperture corresponds to $A=1$ inside the opening and $A=0$ outside.  
In practice, mild edge smoothing of $A(x,y)$ improves numerical stability by reducing high-frequency artifacts.

This paraxial Fourier propagation scheme reduces diffraction to just two 2D FFTs per propagation step.  
It is particularly well suited for keV electron optics: efficient, strictly unitary, accurate in the small-angle regime, and readily extendable to multi-aperture geometries. For independent verification, scripts and reference output images are available online~\cite{NaumovGitHub}.

\subsection*{Numerical rendering and experimental scaling.}

All panels in this work display the \textit{expected counts per detector bin}, rather than a dimensionless probability density.
In the simulations, the incident flux through the aperture is normalized to unity, so that the total signal on the screen integrates to the detector acceptance \(F_{\rm det}\le 1\) within the chosen field of view.
Let \(P_{ij}\) denote the fraction of the total transmitted flux collected by a detector bin of area \(\Delta A=\Delta x\,\Delta y\), such that \(\sum_{ij}P_{ij}=F_{\rm det}\).

For a pulsed source with bunch (pulse) charge \(Q\) and repetition rate \(f_{\rm rep}\), the average beam current is
\begin{equation}
I_{\rm beam} \;=\; Q\,f_{\rm rep},
\label{eq:Ibeam}
\end{equation}
The corresponding particle rate is \(I_{\rm beam}/e\) for singly charged particles, or \(I_{\rm beam}/(Z e)\) for ions of charge state \(Z\), where \(e\simeq1.602\times10^{-19}\,\mathrm{C}\) is the elementary charge.
For a detector with quantum efficiency \(\eta\) and exposure time \(\Delta t\), the expected number of detected events in bin \((i,j)\) is
\begin{equation}
N_{\rm exp}(i,j) \;=\; \eta\,\frac{I_{\rm beam}}{e}\,\Delta t\,P_{ij}.
\label{eq:Nexp}
\end{equation}

The quantity \(R_{\rm tot}\) thus represents the \textit{total detected count rate} (counts per second) integrated over all detector bins within the field of view.
\begin{equation}
    R_{\rm tot}=\eta\,\frac{I_{\rm beam}}{e}\,F_{\rm det}
=\eta\,\frac{Q\,f_{\rm rep}}{e}\,F_{\rm det}.
\label{eq:Rtot}
\end{equation}

The accumulation time required to register \(N\) total detected events is
\begin{equation}
    t_N=\frac{N}{R_{\rm tot}}
=\frac{N\,e}{\eta\,F_{\rm det}\,Q\,f_{\rm rep}}.
\label{eq:tN}
\end{equation}

As a concrete reference, for \(Q=1~\mathrm{pC}=10^{-12}~\mathrm{C}\), \(f_{\rm rep}=1~\mathrm{Hz}\), and \(F_{\rm det}=10^{-4}\),
one obtains an average beam current \(I_{\rm beam}=1~\mathrm{pA}\) and a particle rate \(I_{\rm beam}/e\simeq6.24\times10^{6}~\mathrm{s^{-1}}\).
For detector quantum efficiencies \(\eta=0.2\text{--}0.15\), the total count rate is \(R_{\rm tot}\simeq125\text{--}94~\mathrm{s^{-1}}\),
and the time to accumulate \(N=10^{5}\) detected events is
\[
t_{10^{5}} \approx 13\text{--}18~\mathrm{min}.
\]
The detector-plane coordinates \(X,Y\) in all figures are given in micrometers.
Rescaling to other exposure times, efficiencies, or repetition rates follows linearly from the relations above.

\section{Results}\label{sec:results}

\subsection{Circular aperture: symmetry benchmark}\label{subsec:circ}
\begin{figure}[t]
\centering
\includegraphics[width=\linewidth]{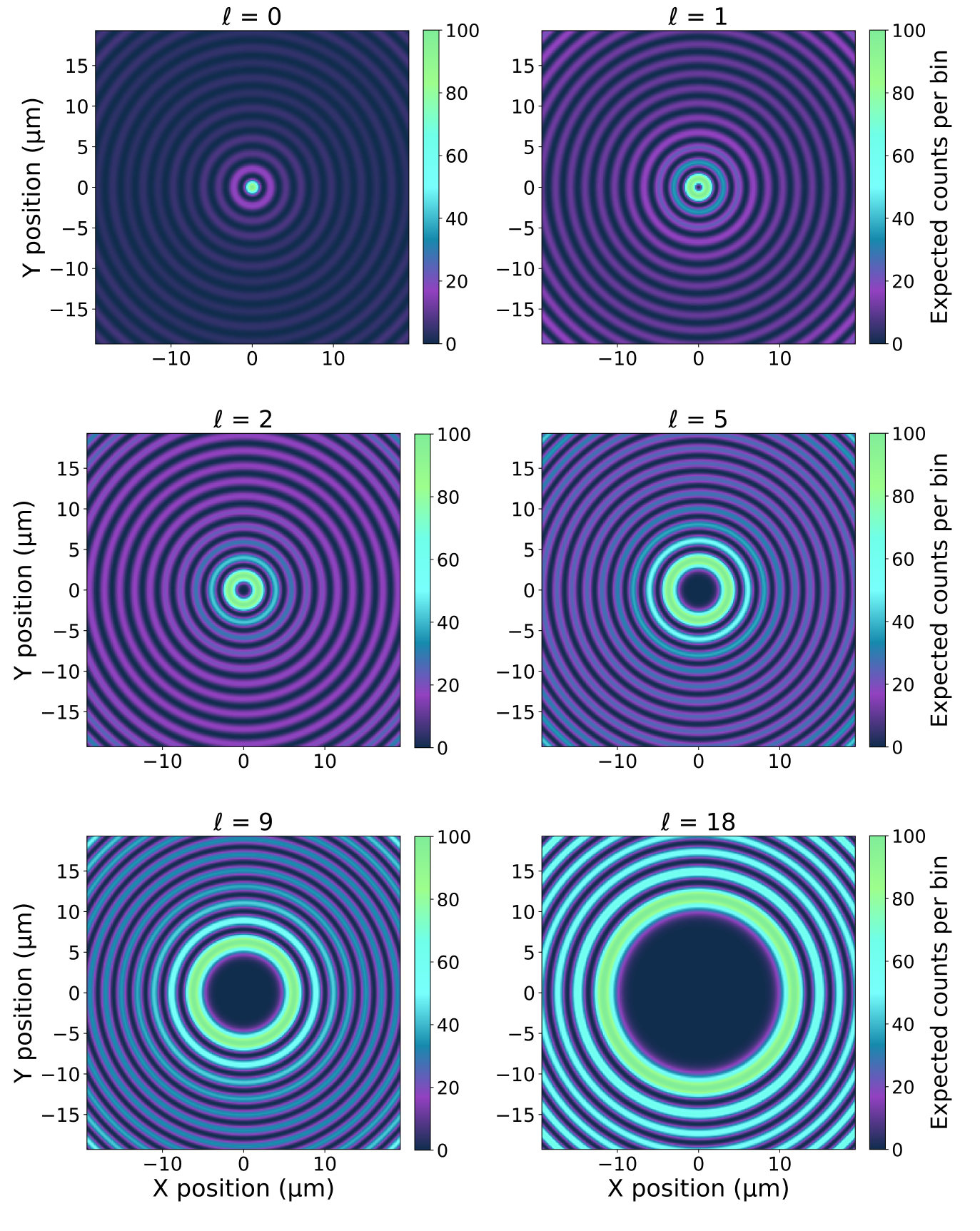}
\caption{%
Circular-aperture benchmark at $E_{\rm kin}=100$~keV, $\kappa=15$~eV.
Expected-count maps for twisted electron Bessel beams with $\ell=0,1,2,5,9,18$ transmitted by a circular aperture of radius $a=400$~nm; aperture–screen distance $z=0.4$~m.
Axial symmetry implies insensitivity to $\mathrm{sign}(\ell)$; the radius of the first bright ring increases with $|\ell|$.
}
\label{fig:bessel_100keV}
\end{figure}

\begin{figure}[t]
\centering
\includegraphics[width=\linewidth]{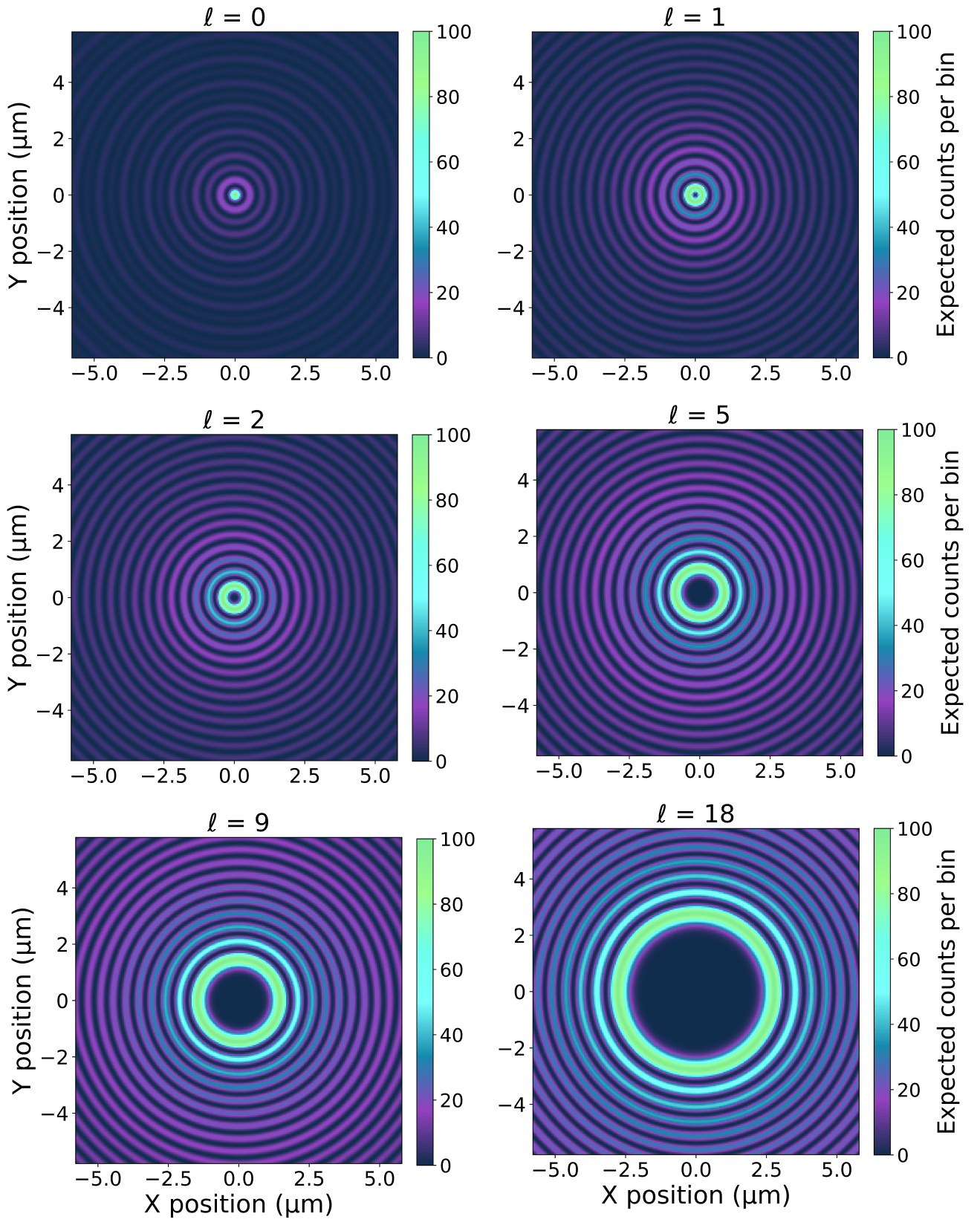}
\caption{%
The same setup as Fig.~\ref{fig:bessel_100keV} but for $E_{\rm kin}=1$~MeV, $\kappa=70$~eV.
The shorter de~Broglie wavelength yields a tighter radial scale and more closely spaced rings, reducing the overall field of view on the detector.
}
\label{fig:bessel_1MeV}
\end{figure}

As a symmetry benchmark, we compute far-field diffraction of twisted electron Bessel beams by a circular aperture of radius $a$ at $z=0$. Owing to axial symmetry, the patterns are insensitive to the sign of the OAM: $\ell$ and $-\ell$ produce identical maps. The on-axis intensity vanishes for $\ell\neq 0$ and is finite only for the fundamental mode. Increasing $|\ell|$ widens the central vortex core and shifts the first bright ring to larger radii, while additional rings arise from radial oscillations of the input profile; see Fig.~\ref{fig:bessel_100keV}. Energy affects the overall radial scale only via the de~Broglie wavelength: in our geometry, raising the kinetic energy from $100$~keV to $1$~MeV (shorter $\lambda_{\rm dB}$) compresses the pattern—compare Fig.~\ref{fig:bessel_100keV} to Fig.~\ref{fig:bessel_1MeV}.

\subsection{Triangular aperture: electrons}\label{subsec:triangle-electrons}
We now turn to diffraction by an equilateral triangular aperture, which breaks axial symmetry and reduces continuous rotations to the discrete group $C_3$ (Sec.~\ref{subsec:triangle-analysis}). In the far field the patterns become OAM-resolving: along each side the number of bright lobes equals $(|\ell|+1)$, and the global orientation flips under $\ell\!\to\!-\ell$.

\begin{figure}[t]
\centering
\includegraphics[width=\linewidth]{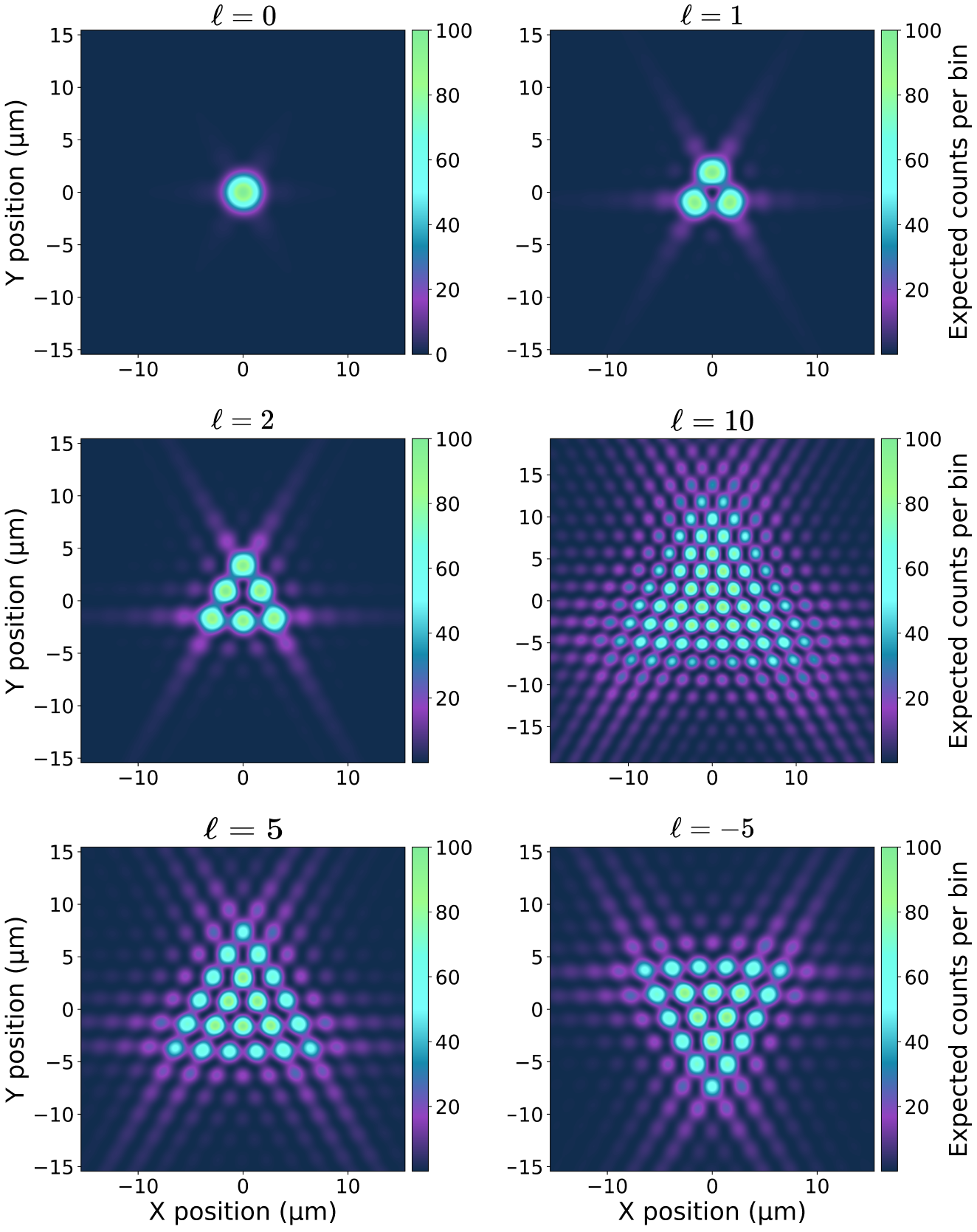}
\caption{%
Triangular-aperture benchmark at $E_{\rm kin}=100$~keV, $\kappa=15$~eV.
Expected-count maps for twisted Bessel electron beams with $\ell=0,1,2,5,-5,10$ transmitted by an equilateral triangular aperture of side $L=400$~nm; aperture–screen distance $z=0.2$~m.
The $(|\ell|+1)$-lobe count per side and the orientation flip under $\ell\!\to\!-\ell$ provide a direct readout of both the magnitude and the sign of the OAM.
}
\label{fig:triangle_bessel_100keV}
\end{figure}

\begin{figure}[t]
\centering
\includegraphics[width=\linewidth]{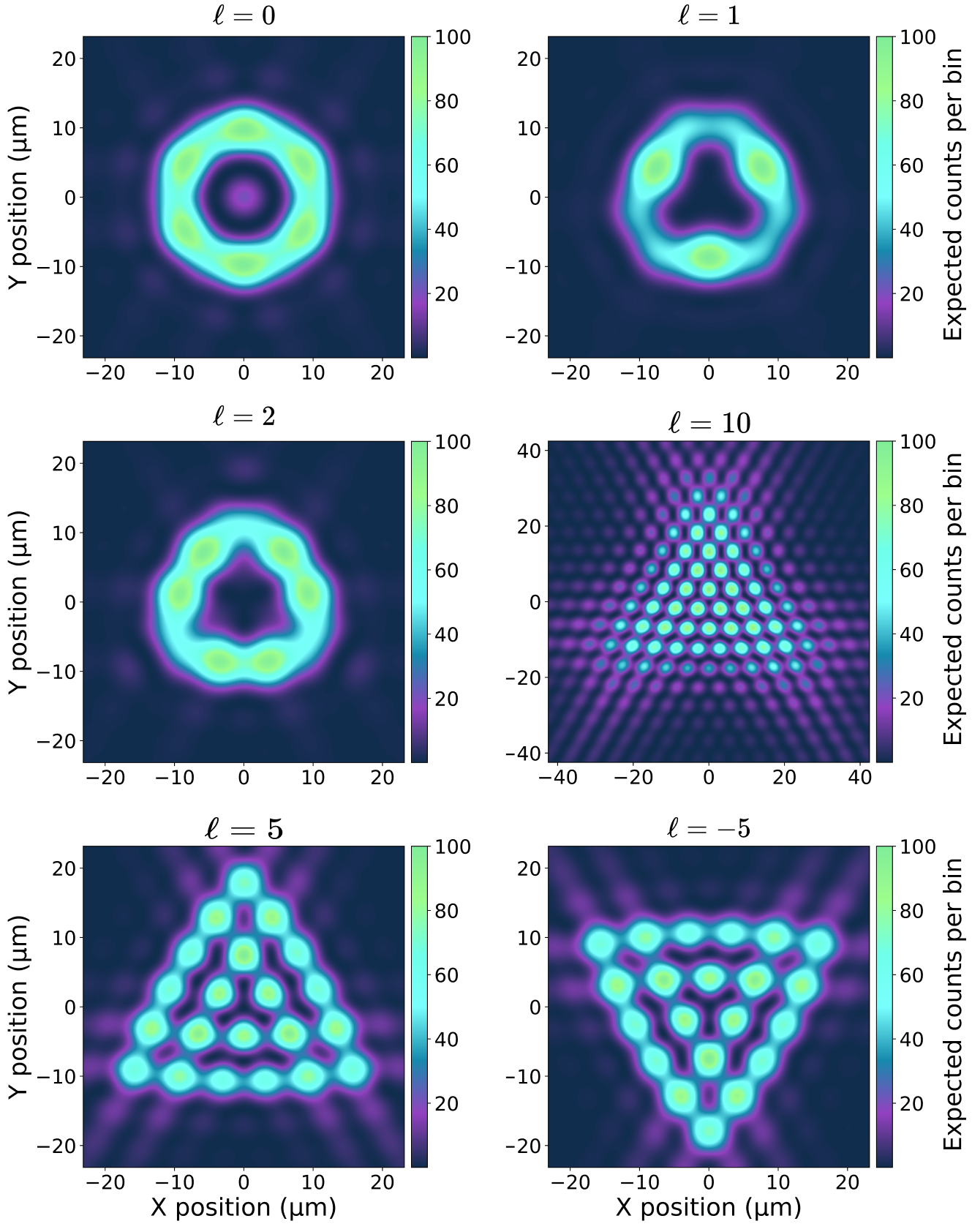}
\caption{%
The same geometry as Fig.~\ref{fig:triangle_bessel_100keV} but for $E_{\rm kin}=1$~MeV, $\kappa=70$~eV and $z=2$~m; triangular side length $L=400$~nm.
The shorter de~Broglie wavelength yields finer fringe spacing, while the longer propagation distance magnifies the triangular pattern on the detector.
}
\label{fig:triangle_bessel_1MeV}
\end{figure}

\begin{figure}[t]
\centering
\includegraphics[width=\linewidth]{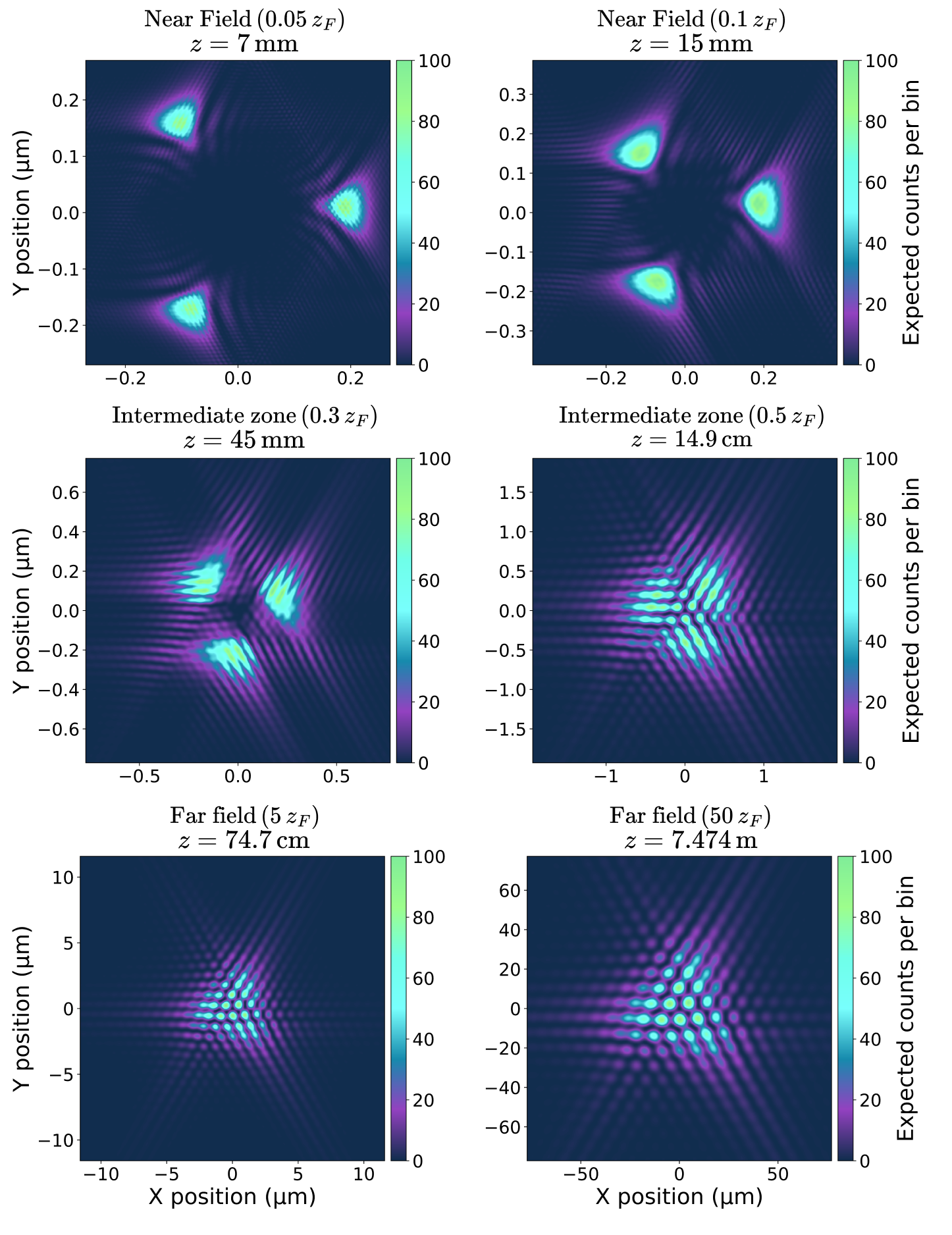}
\caption{%
Near-to-far-field evolution for LG electron wave packet with $\ell=5$ and initial transverse width $\sigma_{\perp}(0)=10~\mathrm{nm}$ diffracted by an equilateral triangular aperture of side $L=400$~nm at $E_{\rm kin}=3$~MeV, shown for increasing screen distances $z$.
The triangular lobe structure sharpens with $z$ and approaches its Fraunhofer form for $z\gtrsim 300$~mm, with $(|\ell|+1)$ bright lobes per side.
}
\label{fig:z_scan_3keV}
\end{figure}

The examples in Figs.~\ref{fig:triangle_bessel_100keV}–-\ref{fig:LG_tri_1MeV} illustrate robust scaling: the far-field triangle grows linearly with $z$, whereas the fringe spacing tightens as the de~Broglie wavelength decreases with energy. For relativistic electrons (e.g., $E_{\rm kin}\!\sim\!1$~MeV), maintaining a $\sim 10\times10~\mu$m field of view simply requires a proportionally longer propagation distance (Sec.~\ref{sec:feasibility} and Table~\ref{tab:design_geometries}).

\subsection{Robustness for LG packets (Bessel vs.\ LG)}\label{subsec:bessel-vs-lg}
\begin{figure}[t]
\centering
\includegraphics[width=\linewidth]{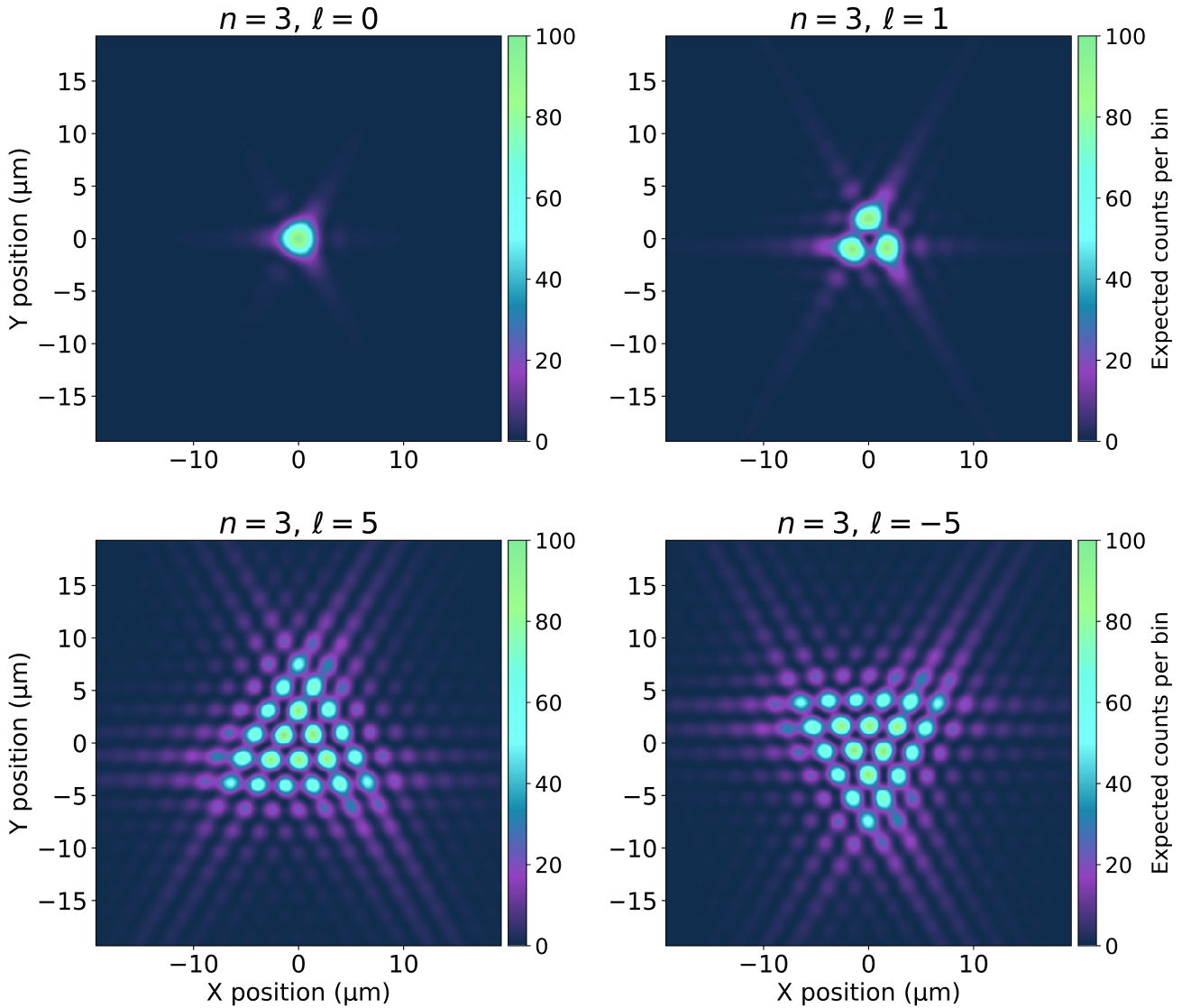}
\caption{%
Triangular-aperture diffraction for LG electron wave packets at $E_{\rm kin}=100$~keV.
Expected-count maps for $n=3$ with $\ell=0,1,5,-5$, transmitted by an equilateral triangular aperture of side $L=400$~nm; aperture--screen distance $z=0.2$~m; LG initial transverse width $\sigma_{\perp}(0)=10~\mathrm{nm}$.
The $|\ell|+1$-lobe count per side and the orientation flip under $\ell\!\to\!-\ell$ match the Bessel-beam results in Fig.~\ref{fig:triangle_bessel_100keV}.
}
\label{fig:LG_tri_100keV}
\end{figure}

\begin{figure}[t]
\centering
\includegraphics[width=\linewidth]{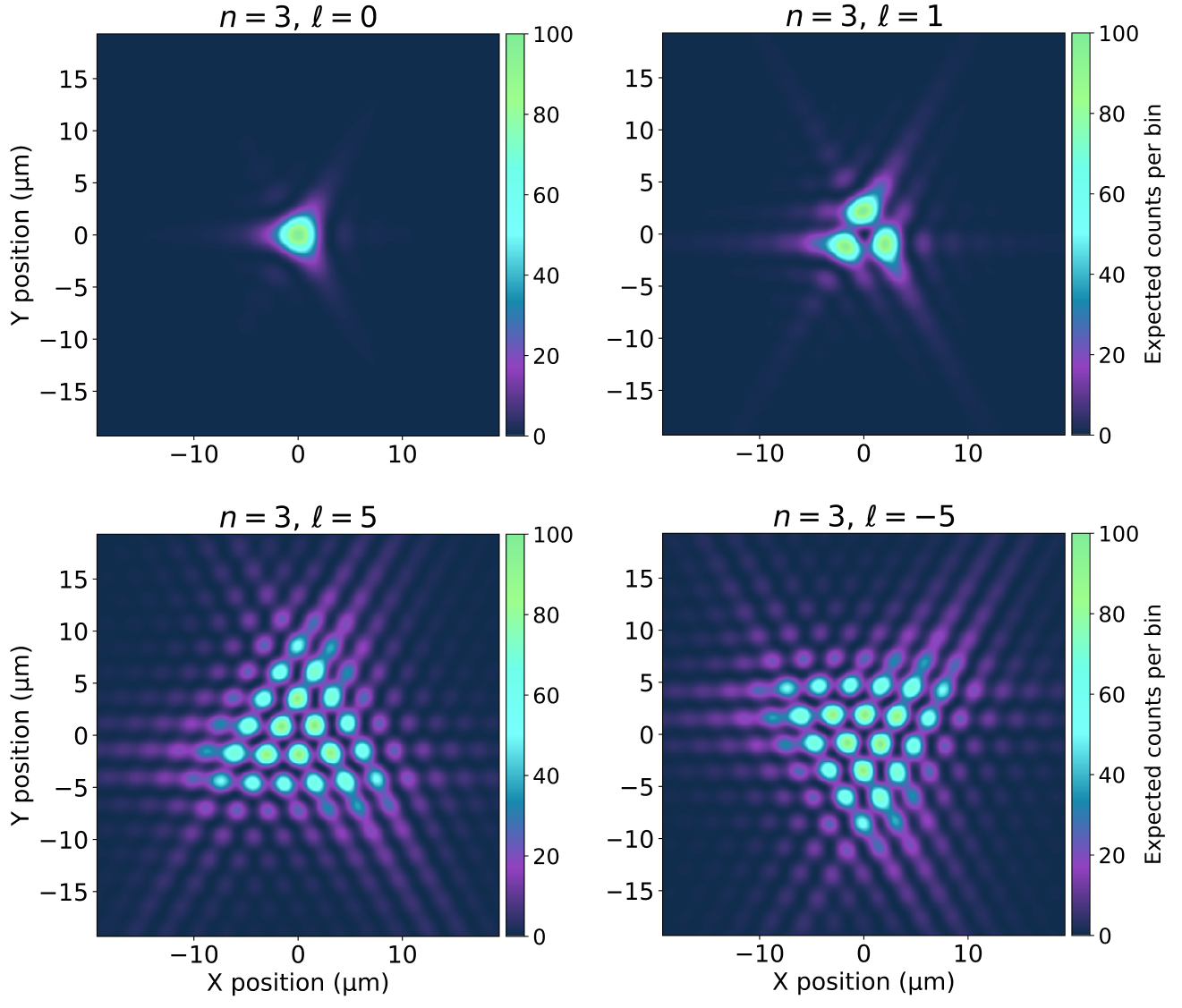}
\caption{%
The same comparison at $E_{\rm kin}=1$~MeV and $z=1$~m.
The reduced de~Broglie wavelength tightens fringe spacing, while the OAM-resolving triangular signature remains nearly identical to the Bessel case in Fig.~\ref{fig:triangle_bessel_1MeV}.
}
\label{fig:LG_tri_1MeV}
\end{figure}

We compare triangular-aperture diffraction for Bessel and LG inputs matched in OAM $\ell$ and in transverse scale at the aperture (Sec.~\ref{subsec:incident-beams}). Within our sampling and dynamic range, the two families produce nearly identical far-field patterns: the OAM-resolving features established for triangular masks—lobe count, handedness, and nodal geometry—remain the same. Differences are confined to the overall envelope and contrast: the LG radial index $n$ modifies the envelope via $L_n^{|\ell|}$ and the width via $M=2n+|\ell|+1$, affecting primarily contrast but neither the $(|\ell|+1)$ lobe count nor the $\mathrm{sign}(\ell)$ sensitivity. For circular apertures (Sec.~\ref{subsec:circ}), the two models likewise yield nearly identical ring patterns under the same matching (App.~\ref{app:lg-bessel}). Representative diffraction maps for LG inputs are shown in Figs.~\ref{fig:LG_tri_100keV} and~\ref{fig:LG_tri_1MeV}.

\subsection{Twisted ions}\label{subsec:ions}
We now evaluate triangular-aperture diffraction for twisted ions. Compared with typical electron-beam conditions, realistic ion sources provide very small initial transverse widths $\sigma_\perp(0)$ and short de~Broglie wavelengths, which yield short Rayleigh lengths. Localized Bessel profiles are therefore not preserved over centimeter drifts. We model the incident state as a LG packet (Sec.~\ref{subsec:incident-beams}).

\paragraph*{Modeling and geometry.}
To enable species-to-species comparison we use a common layout: source–aperture distance $d_{\rm sa}=10~\mathrm{cm}$, equilateral triangular opening of side $L=40~\mathrm{nm}$ at $z=0$, and aperture–screen distance $z=2~\mathrm{m}$. For the ion energies considered here this places the detector well in the far field (Sec.~\ref{sec:feasibility}). The boundary field at the mask is the LG packet evaluated at $z=d_{\rm sa}$.

\begin{figure}[t]
\centering
\includegraphics[width=\linewidth]{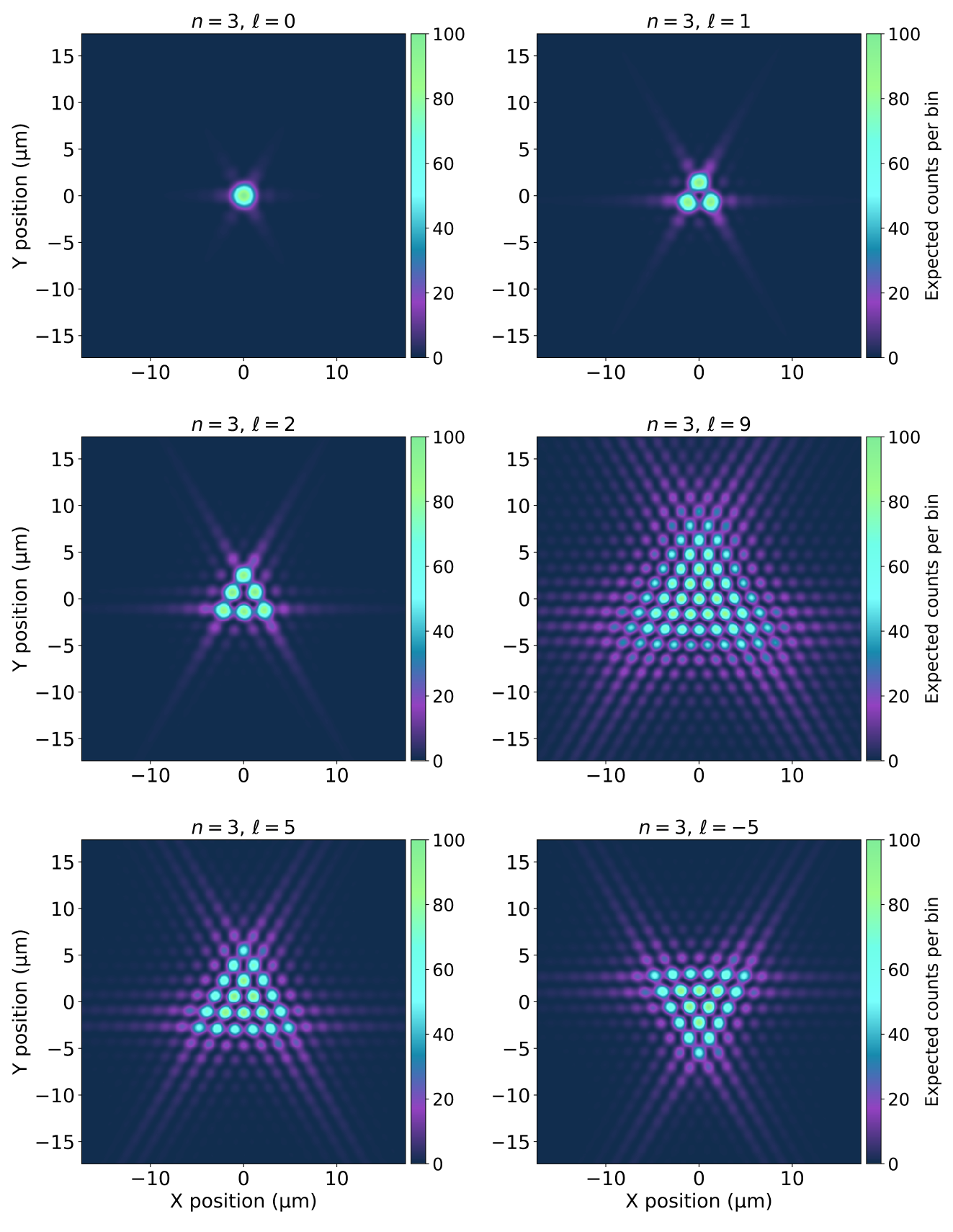} 
\caption{%
Triangular-aperture diffraction for protons modeled as LG packets ($n=3$; $\ell=0,1,5,-5,9$) transmitted by an equilateral triangular aperture of side $L=40~\mathrm{nm}$. Beam kinetic energy $E_{\rm kin}=1~\mathrm{MeV}/u$; initial transverse width $\sigma_\perp(0)=10~\mathrm{pm}$; source–aperture distance $d_{\rm sa}=10~\mathrm{cm}$; aperture–screen distance $z=2~\mathrm{m}$. The OAM-resolving triangular lattice is clearly visible.%
}
\label{fig:ion_proton_1MeV}
\end{figure}

\begin{figure}[t]
\centering
\includegraphics[width=\linewidth]{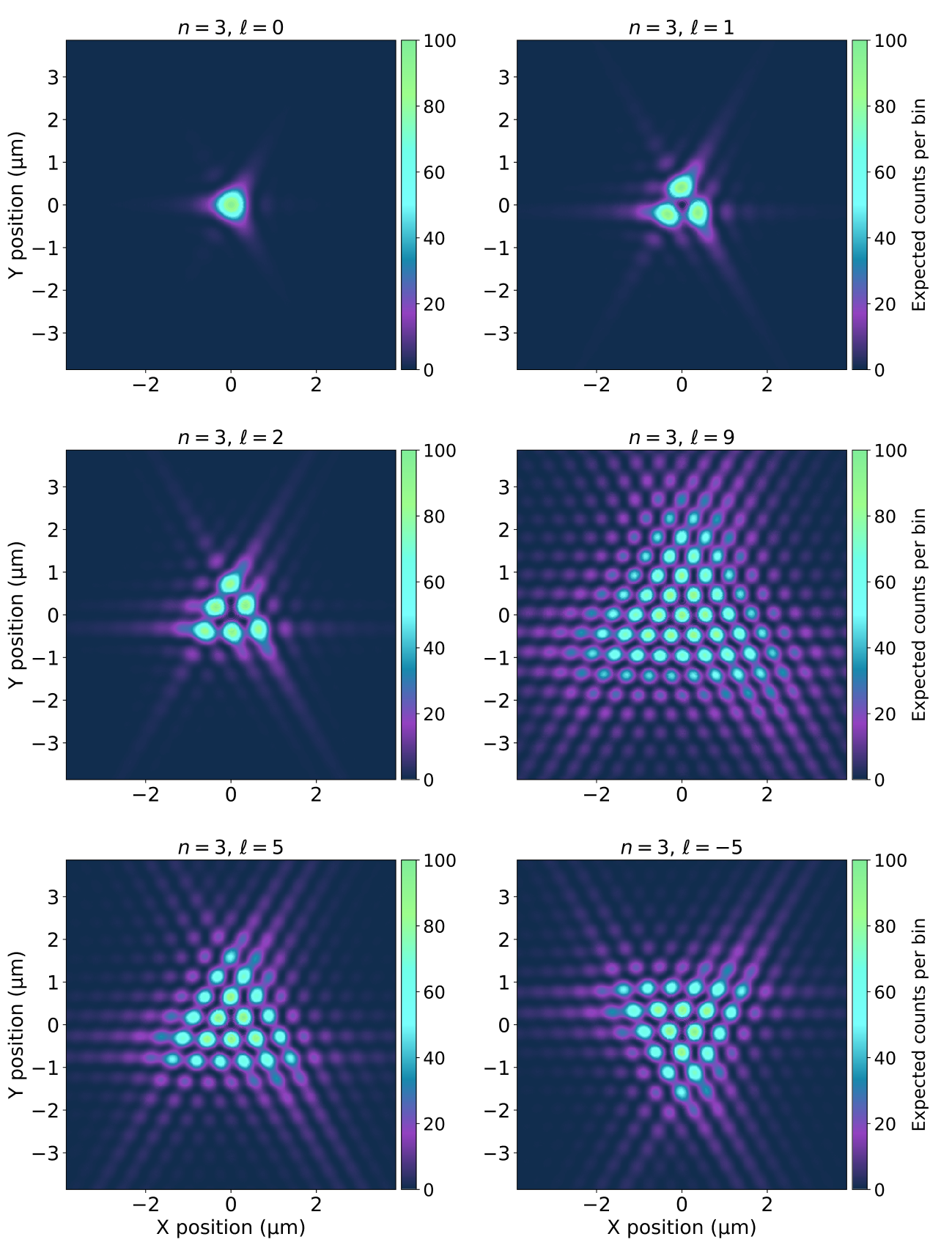} 
\caption{%
The same geometry and LG modeling as Fig.~\ref{fig:ion_proton_1MeV}, but for ${}^{12}\mathrm{C}^{6+}$ at $E_{\rm kin}=1~\mathrm{MeV}/u$ and the same $\sigma_\perp(0)$. The OAM-resolving triangular signature is preserved, while the overall image contracts to $\sim 3\times 3~\mu\mathrm{m}^2$.%
}
\label{fig:ion_carbon_1MeV}
\end{figure}

\paragraph*{Triangular aperture: protons vs.\ carbon.}
Figure~\ref{fig:ion_proton_1MeV} shows expected-count maps for protons at $E_{\rm kin}=1~\mathrm{MeV}/u$ with $\sigma_\perp(0)=10~\mathrm{pm}$ and $n=3$ for several $\ell$. The hallmark OAM features are evident: along each side the number of bright lobes equals $(|\ell|+1)$ and the global orientation flips under $\ell\!\to\!-\ell$. For the same geometry, ${}^{12}\mathrm{C}^{6+}$ at $E_{\rm kin}=1~\mathrm{MeV}/u$ produces a similar OAM-resolving triangular lattice with a smaller overall field of view (Fig.~\ref{fig:ion_carbon_1MeV}), reflecting the shorter de~Broglie wavelength at fixed $z$ and $L$. The handedness and lobe count are unchanged. Quantitative layouts and detector requirements for multi-MeV ions are collected in Sec.~\ref{sec:feasibility}.

\section{Design rules and feasibility}\label{sec:feasibility}
We collect here practical rules and scalings used throughout the paper, with pointers to the theory and examples in Secs.~\ref{subsec:circ}--\ref{subsec:ions}. All image panels show expected counts normalized as in Sec.~\ref{sec:numerics}.

\paragraph*{Far-field criterion.}
The Fresnel--Fraunhofer crossover is set by Eq.~\eqref{eq:fraunhofer}. For an equilateral triangle we take $D\simeq L/\!\sqrt{3}$. For instance, with $L=0.5~\mu$m and electrons of $E_{\rm kin}=0.1$--$5$~MeV, Eq.~\eqref{eq:fraunhofer} yields $z_{\mathrm F}\sim 20$--$400$~mm; beyond this range the pattern is governed primarily by aperture geometry and by the beam’s angular spectrum.

\paragraph*{Image scale, lattice pitch, and detector sampling.}
In the Fraunhofer regime detector coordinates map linearly to the Fourier plane [Eq.~\eqref{eq:XY-from-k}], and the three ridge families of the triangular spectrum generate a regular detector lattice with pitch $\Delta$ from Eq.~\eqref{eq:Delta-step}. Increasing $z$ magnifies the image linearly, while increasing $L$ compresses it. Across the geometries in Secs.~\ref{subsec:circ}--\ref{subsec:ions}, salient features lie in the $1$--$10~\mu$m range. For good contrast we require detector pixel pitch $p\!\lesssim\!\Delta/3$; as $p\!\to\!\Delta/2$, lobe visibility and handedness readout degrade rapidly.

\paragraph*{Optimization under detector constraints.}
Fix a target lattice pitch $\Delta$ (set by sampling via Eq.~\eqref{eq:Delta-step}) and impose a safety margin $C$ in $z\!\ge\! C D^2/\lambda_{\rm dB}$ [Eq.~\eqref{eq:fraunhofer} with $D\simeq L/\!\sqrt{3}$]. Maximizing brightness at fixed $\Delta$ gives
\begin{equation}
L_{\rm opt}=\frac{\sqrt{3}}{2\,C}\,\Delta,\qquad
z_{\rm opt}=\frac{3}{4\,C}\,\frac{\Delta^{2}}{\lambda_{\rm dB}}.
\label{eq:opt-Lz}
\end{equation}
At these values $I_{\rm opt}\!\propto\!3/(4C^{2})$, i.e., independent of $\lambda_{\rm dB}$ at fixed $\Delta$ (derivation in App.~\ref{app:triangle-derivations}). 
Tables~\ref{tab:opt-electrons} and~\ref{tab:opt-protons} list representative optima for $C=10$.

\begin{table}[t]
\caption{Optimized triangular-aperture geometry for electrons with different $E_{\rm kin}$ and $C=10$.
Here $\Delta$ is the detector-lattice pitch set by sampling, $L_{\rm opt}$ is the corresponding optimal side of the triangular aperture, and $z_{\rm opt}$ is the required aperture--detector distance.}
\label{tab:opt-electrons}
\centering
\setlength{\tabcolsep}{3pt}
\renewcommand{\arraystretch}{1.05}
\begin{ruledtabular}
\begin{tabular}{ccc}
$\Delta$ ($\mu$m) & $L_{\rm opt}$ (nm) & $z_{\rm opt}$ (m) for $E_{\rm kin}$ 0.1/1/3 MeV \\
\hline
1.00 & 86.60  & 0.02 / 0.09 / 0.21 \\
2.50 & 216.51 & 0.13 / 0.54 / 1.31 \\
5.00 & 433.01 & 0.51 / 2.15 / 5.25 \\
\end{tabular}
\end{ruledtabular}
\end{table}

\begin{table}[t]
\caption{Optimized triangular-aperture geometry for protons with $E_{\rm kin}=1$~MeV and $C=10$.
Here $\Delta$ is the detector-lattice pitch set by sampling, $L_{\rm opt}$ is the corresponding optimal side of the triangular aperture, and $z_{\rm opt}$ is the required aperture--detector distance.}
\label{tab:opt-protons}
\centering
\setlength{\tabcolsep}{5pt}
\renewcommand{\arraystretch}{1.05}
\begin{ruledtabular}
\begin{tabular}{ccc}
$\Delta$ ($\mu$m) & $L_{\rm opt}$ (nm) & $z_{\rm opt}$ (m) \\
\hline
0.50 & 43.30  & 0.66 \\
1.00 & 86.60  & 2.62 \\
2.00 & 173.21 & 10.49 \\
\end{tabular}
\end{ruledtabular}
\end{table}

These values highlight the trade-off: at fixed $\Delta$ the optimal side $L_{\rm opt}$ is set purely by $C$, whereas $z_{\rm opt}$ increases with energy (smaller $\lambda_{\rm dB}$). For the electron cases in Table~\ref{tab:opt-electrons} with $\Delta=1$--$5~\mu$m one has $L_{\rm opt}\!\approx\!0.09$--$0.43~\mu$m and $z_{\rm opt}$ ranging from a few centimeters up to $\lesssim 1$~m at $0.1$~MeV, to the meter scale at $1$~MeV, and to several meters at $3$~MeV. For protons at $1$~MeV (Table~\ref{tab:opt-protons}), $\Delta=0.5$--$2~\mu$m implies $L_{\rm opt}\!\approx\!0.04$--$0.17~\mu$m and $z_{\rm opt}$ from sub-meter to $\sim 10$~m, i.e., meter-scale drifts are typical; heavier ions push $z_{\rm opt}$ further upward for the same $\Delta$.

\paragraph*{Scaling to multi-MeV electrons.}
While the examples in Sec.~\ref{subsec:triangle-electrons} focus on the $0.1$–$1$~MeV range relevant to electron microscopy, the triangular-aperture method extends into the multi-MeV domain accessible at accelerator beamlines. As summarized in Table~\ref{tab:design_geometries}, for $E_{\rm kin}=3$~MeV the method still yields a well-resolved triangular signature with a $\sim10\times10~\mu\mathrm m^2$ field of view at $z=2$~m. At higher energies the minimal propagation distance required to maintain this image scale grows approximately linearly with electron energy at fixed initial packet width. For instance, at $E_{\rm kin}=5$~MeV one preserves a $\sim10\times10~\mu\mathrm m^2$ image only by increasing both the source–aperture and aperture–detector distances to several meters (e.g., $d_{\rm sa}\!\sim\!0.6$~m and $z\!\sim\!4$~m). Alternatively, at shorter distances (e.g., $d_{\rm sa}\!\sim\!0.3$~m and $z\!\sim\!1$~m) the same beam produces a triangular pattern reduced to $\sim3\times3~\mu\mathrm m^2$. Thus the method remains viable up to a few MeV, but the shrinking $\Delta$ and increasing drift lengths tighten layout constraints at ultra\-relativistic energies.

\paragraph*{Ions: packet spreading and image size.}
For realistic ion sources the attainable initial transverse width $\sigma_\perp(0)$ is very small and the de~Broglie wavelength is short, resulting in short Rayleigh lengths. The spatiotemporal evolution discussed in Sec.~\ref{subsec:incident-beams} implies that localized Bessel-like profiles are not preserved over centimeter drifts; LG modeling is appropriate.
For proton beams with $E_{\rm kin}=1$~MeV and $\sigma_\perp(0)\!\sim\!1$--$10$~pm one finds $z_R\!\sim\!0.2$--$20$~nm, so spreading is significant.
For a proton at $E_{\rm kin}=1$~MeV with initial transverse coherence $\sqrt{\langle\rho^2\rangle}(0)=1$--$10$~pm, the registered width in the observation plane increases to $\sim100~\mu$m after propagation of $\sim0.02$--$0.2$~m.
For ${}^{12}\mathrm C^{6+}$ at the same $E_{\rm kin}$ and $\sqrt{\langle\rho^2\rangle}(0)$, the same width $\sim100~\mu$m is reached after $\sim0.08$--$0.8$~m, reflecting the longer Rayleigh length of the heavier ion.
In the common layout used here (e.g., $d_{\rm sa}=10$~cm, $z=2$~m, $L=40$~nm) both proton and carbon cases lie comfortably in the far field; the proton image spans $\sim10\times10~\mu\mathrm m^2$, while ${}^{12}\mathrm C^{6+}$ contracts to $\sim3\times3~\mu\mathrm m^2$. Taking $L=200$~nm with $d_{\rm sa}\!\approx\!1$~m and $z\!\approx\!2$~m for 1~MeV protons provides the field of view $\sim 1\times1~\mu\mathrm m^2$ (Table~\ref{tab:design_geometries}).

\begin{table*}[t]
\caption{Representative geometries and expected image scales.
Columns: Species; $E_{\rm kin}$ — kinetic energy for electrons, energy per nucleon for ions; $\sigma_\perp(0)$ — initial transverse rms; $L$ — triangle side; $d_{\rm sa},\,z$ — source--aperture distance in cm and aperture--detector distance in m; $\Delta$ — lattice pitch from Eq.~\eqref{eq:Delta-step}; Image size — field of view; $p_{\min}$ — recommended pixel pitch ($p_{\min}\!\lesssim\!\Delta/3$).}
\label{tab:design_geometries}
\squeezetable
\begin{ruledtabular}
\begin{tabular*}{\textwidth}{@{\extracolsep{\fill}}lcccccc}
Species & $E_{\rm kin}$/\,$\sigma_\perp(0)$ & $L$ (nm) & $(d_{\rm sa},z)$ (cm, m) & Image size ($\mu$m$^2$) & $\Delta$ ($\mu$m) & $p_{\min}$ ($\mu$m) \\
\hline
H$^+$           & $1000/10$   & $40$  & $(10.0,\,2.0)$  & $\sim 10\times 10$ & {\small $\sim 2.0$}  & {\small $\lesssim 0.67$} \\
H$^+$           & $1000/10$   & $200$ & $(100.0,\,2.0)$ & $\sim 1\times 1$   & {\small $\sim 0.2$}  & {\small $\lesssim 0.07$} \\
$^{12}$\!C$^{6+}$ & $1000/10$ & $40$  & $(10.0,\,2.0)$  & $\sim 3\times 3$   & {\small $\sim 0.6$}  & {\small $\lesssim 0.2$} \\
e$^-$ (non-rel.)& $100/10^4$  & $400$ & $(4.0,\,0.2)$   & $\sim 10\times 10$ & {\small $\sim 2.0$}  & {\small $\lesssim 0.67$} \\
e$^-$ (rel.)    & $1000/10^4$ & $400$ & $(8.0,\,1.0)$   & $\sim 10\times 10$ & {\small $\sim 2.0$}  & {\small $\lesssim 0.67$} \\
e$^-$ (rel.)    & $3000/10^4$ & $400$ & $(15.0,\,2.0)$  & $\sim 10\times 10$ & {\small $\sim 2.0$}  & {\small $\lesssim 0.67$} \\
\end{tabular*}
\end{ruledtabular}
\end{table*}

\section{Discussion}\label{sec:discussion}
Our results demonstrate that the geometry of an aperture is the key factor determining whether diffraction is sensitive to the OAM of twisted matter waves. Circular apertures preserve axial symmetry and produce ring-like far-field patterns that are insensitive to the OAM sign and only weakly dependent on $|\ell|$. Such configurations are useful for characterizing beam collimation and radial coherence, but they cannot unambiguously resolve different OAM states. In contrast, triangular apertures break rotational symmetry and generate structured diffraction patterns whose lobe number and orientation provide a direct signature of both $|\ell|$ and $\mathrm{sign}(\ell)$. This behavior follows from symmetry arguments and persists across particle types, energies, and propagation distances. Importantly, the triangular OAM sensitivity is robust not only for ideal Bessel inputs but also for localized Laguerre--Gaussian wave packets that undergo transverse spreading. Direct numerical simulations using the split--step Fourier method confirm the Kirchhoff-based predictions within numerical accuracy.

From a practical perspective, two experimental limitations arise. First, for relativistic electrons with energies above $\sim 10$\,MeV, the reduced de~Broglie wavelength requires propagation distances exceeding several meters to reach the Fraunhofer regime. At such distances the triangular diffraction pattern contracts to only a few micrometers across (e.g., $\sim 5\times 5~\mu$m at $z\gtrsim 5$\,m), which renders a direct implementation with a planar mask experimentally very challenging. One possible route to relax the strict far-field requirement is to use a \textit{controlled curvature} of the screening surface: a \textit{concave} triangular mask oriented toward the detector can partially compensate path-length differences in the intermediate (pre-wave) zone, thereby enhancing image contrast and reducing spot size. Similar concepts are well established in electromagnetic diffraction on curved surfaces~\cite{Potylitsyn2011}. The practical feasibility of fabricating such curved masks at the few-hundred-nanometer scale is challenging, but this strategy offers a potential route to mitigate long propagation distances.

Second, for ions the situation is constrained by both aperture size and vacuum requirements. Already for $^{12}\mathrm{C}^{6+}$ ions at $E_{\rm kin}=1$\,MeV/u and $z=2$\,m the diffraction image contracts to $\sim 3\times 3~\mu$m, while the aperture size must be kept at the tens-of-nanometers scale, pushing current fabrication limits. Moreover, increasing the triangular side well beyond the few-hundred-nanometer range is generally counterproductive: at fixed wavelength and drift distances the detector-plane pitch shrinks as $1/L$, so a tenfold larger aperture yields a tenfold smaller image. To recover a $\sim 10\times10~\mu$m field of view one must then extend both the source–aperture and aperture–screen distances to the $10$-m scale (cf. Sec.~\ref{sec:feasibility}), which can be impractical for some beamlines. These considerations set realistic upper bounds for the applicability of triangular diffraction as an OAM diagnostic tool. Table~\ref{tab:design_geometries} summarizes the key control parameters used in our simulations and the resulting image scale at the detector for representative species.

Finally, the characteristic modulation scale in the calculated diffraction images is on the order of $5$–$10~\mu$m. This is comparable to or smaller than the pixel size of typical silicon detectors ($20$–$50~\mu$m), and therefore experimental verification will require detection systems with an effective spatial resolution better than $1~\mu$m. For non-single-particle electron and ion beams in the energy range from $100~\text{keV}$ up to several $\text{MeV}$, achieving such resolution is challenging due to multiple scattering and charge diffusion in thick detection layers. High-grain photographic emulsions can in principle record micron-scale features, but their performance rapidly degrades at high particle energies. More practical approaches involve microchannel plates (MCPs) with optical or phosphor readout, or scintillating screens combined with electron-optical projection systems providing moderate magnification. Electron lenses routinely employed in PEEM/LEEM and electron holography setups~\cite{schramm2012peemleem,verbeeck2010nature,beche2014magnetic} can magnify the diffraction pattern by factors of $10$–$50$ before detection, effectively relaxing the submicron-resolution requirement while preserving spatial coherence. Such hybrid electron-optical imaging schemes have already been demonstrated for vortex-electron and magnetic-monopole analog experiments. Thus, while the triangular aperture remains a conceptually simple and robust OAM diagnostic, realizing its potential for relativistic electrons and heavy ions will require careful optimization of the beamline geometry, aperture curvature, and imaging system to balance resolution, coherence, and signal intensity.

\section{Conclusions}\label{sec:conclusion}
Although diffraction and interference of matter waves are most often explored in the nonrelativistic regime, we show that triangular-aperture diffraction provides a viable route to diagnose vortex beams well into the moderately relativistic domain. With realistic choices of the aperture side, detector sampling and the boundary of the Fraunhofer zone, the method enables retrieval of OAM for electron beams at the MeV scale, relevant for modern electron guns and small circular accelerators such as microtrons, and for typical ion sources. Required source–aperture and aperture–detector distances can reach a few metres in some cases but remain within technological reach, and may be reduced by employing concave focusing masks or modest electron-optical magnification. Overall, triangular diffraction offers a simple, passive and experimentally practical OAM diagnostic tool for structured quantum beams.

\begin{acknowledgments}
The studies of circular and triangular aperture diffraction presented in Secs.~\ref{sec:theory}--~\ref{sec:results} were supported by the Russian Science Foundation under Grant No.~23-62-10026 \cite{RSF2023}. 
The studies of feasibility and optimization under detector constraints in Secs.~\ref{sec:feasibility}--\ref{sec:conclusion} were supported by the Russian Science Foundation under Grant No.~25-71-00060.
We are grateful to I.~Pavlov, N. Anfimov and A.~Dyatlov for many fruitful discussions and valuable advice. 
\end{acknowledgments}

\appendix

\section{LG vs Bessel: additional details}\label{app:lg-bessel}
This appendix complements Secs.~\ref{subsec:incident-beams}, \ref{subsec:circ}, \ref{subsec:triangle-electrons}, and \ref{subsec:bessel-vs-lg} by recording the boundary fields and normalizations used in our simulations and by providing auxiliary maps not shown in the main text.

\subsection*{Circular aperture: integral and inputs (cylindrical form)}

For a circular aperture of radius $a$ we use cylindrical coordinates $(\rho,\phi,z)$ for the observation point and $(\rho',\varphi',0)$ in the aperture plane. The scalar Kirchhoff–Fresnel integral reads
\begin{align}
\psi(\rho,\phi,z)
&=
\int_{0}^{a}\!\rho'\,d\rho'
\int_{0}^{2\pi}\!d\varphi'\;
\frac{e^{ikR}}{4 \pi R}\,
\Bigg[\Big(ik-\frac{1}{R}\Big)\frac{z}{R}+ik_z\Bigg]
\nonumber\\[1pt]
&\quad\times\,
\psi_{\rm in}(\rho',\varphi',0),
\label{eq:kirchhoff-circ}
\end{align}
where $R=\sqrt{\rho^2+\rho'^2-2\rho\rho'\cos(\phi-\varphi')+z^2}.$
The incident field at the aperture is taken either as an ideal Bessel mode,
\begin{equation}
\psi^{\rm B}_{\ell}(\rho',\varphi',0)
= J_{|\ell|}(\kappa\rho')\,e^{i\ell\varphi'},
\label{eq:bessel-circ}
\end{equation}
or as an LG packet propagated from a source at distance $d_{\rm sa}$,
\begin{align}
\psi^{\rm LG}_{\ell,n}(\rho',\varphi';d_{\rm sa})
&= \mathcal N\;
{\rho'}^{|\ell|}\!
\left[\frac{1}{\sigma_\perp(d_{\rm sa})}\right]^{|\ell|+1}
L^{|\ell|}_n\!\!\left(\frac{\rho'^2}{\sigma_\perp^2(d_{\rm sa})}\right)
e^{i\ell\varphi'} \nonumber\\[1pt]
&\quad\times
\exp\!\left[-\,\frac{\rho'^2}{2\sigma_\perp^2(d_{\rm sa})}
\left(1-\frac{i\,d_{\rm sa}}{z_R}\right)\right]
\nonumber\\[1pt]
&\quad\times
\exp\!\left[-\,i\,M\,\arctan\!\left(\frac{d_{\rm sa}}{z_R}\right)\right],
\label{eq:LG-circ}
\end{align}
We use the standard plane normalization for LG 
\begin{equation}
N^{\mathrm{LG}}_{\ell n}
=\frac{1}{\sigma_\perp(d_{\rm sa})}\,
\sqrt{\frac{n!}{\pi\,(n+|\ell|)!}}\,
\label{eq:app-lg-norm}
\end{equation}
and, when displaying expected-count maps, renormalize on the open set $S$ of the mask so that $\int_S|\psi|^2\,dx'\,dy'=1$.

\subsection*{Triangular aperture: integral and inputs (planar form)}

For a planar screen at $z=0$ pierced by an equilateral triangular opening $S_\triangle$,
\begin{align}
\psi(x,y,z)
&= \int_{S_\triangle}\!
\frac{e^{ikR}}{4\pi R}\,
\Bigg[\Big(ik-\frac{1}{R}\Big)\frac{z}{R}+ik_z\Bigg]
\nonumber\\[1pt]
&\quad\times
\psi_{\rm in}(x',y',0)\;dx'\,dy',
\label{eq:kirchhoff-tri}
\end{align}
where \( R=\sqrt{(x-x')^2+(y-y')^2+z^2} \).
The boundary fields $\psi_{\rm in}$ are the same Bessel and LG forms as above, with
$\rho'=\sqrt{x'^2+y'^2}$ and $\varphi'=\arg(x'+iy')$ in the aperture plane.

\subsection*{Auxiliary maps}

\begin{figure}[t]
\centering
\includegraphics[width=\linewidth]{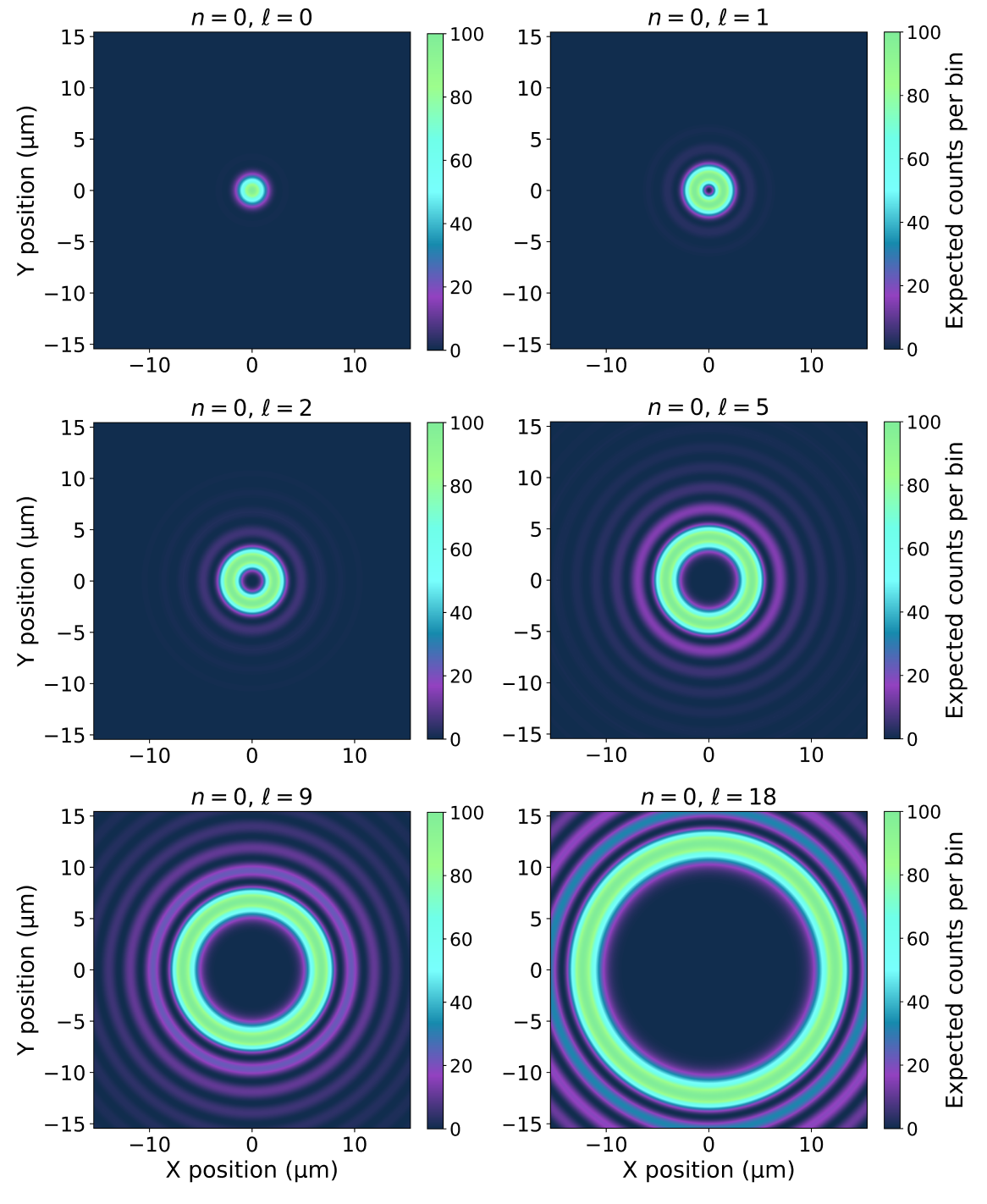}
\caption{LG packets diffracted by a circular aperture (expected-count maps).
$E_{\rm kin}=100$~keV; aperture radius $a=400$~nm; screen distance $z=0.4$~m; initial transverse width $\sigma_{\perp}(0)=10~\mathrm{nm}$.
Shown for $n=0$ and $\ell=0,1,2,5,9,18$. These auxiliary maps serve as a symmetry benchmark (cf.\ Sec.~\ref{subsec:circ}).}
\label{fig:LG-circ-app}
\end{figure}

\begin{figure}[t]
\centering
\includegraphics[width=\linewidth]{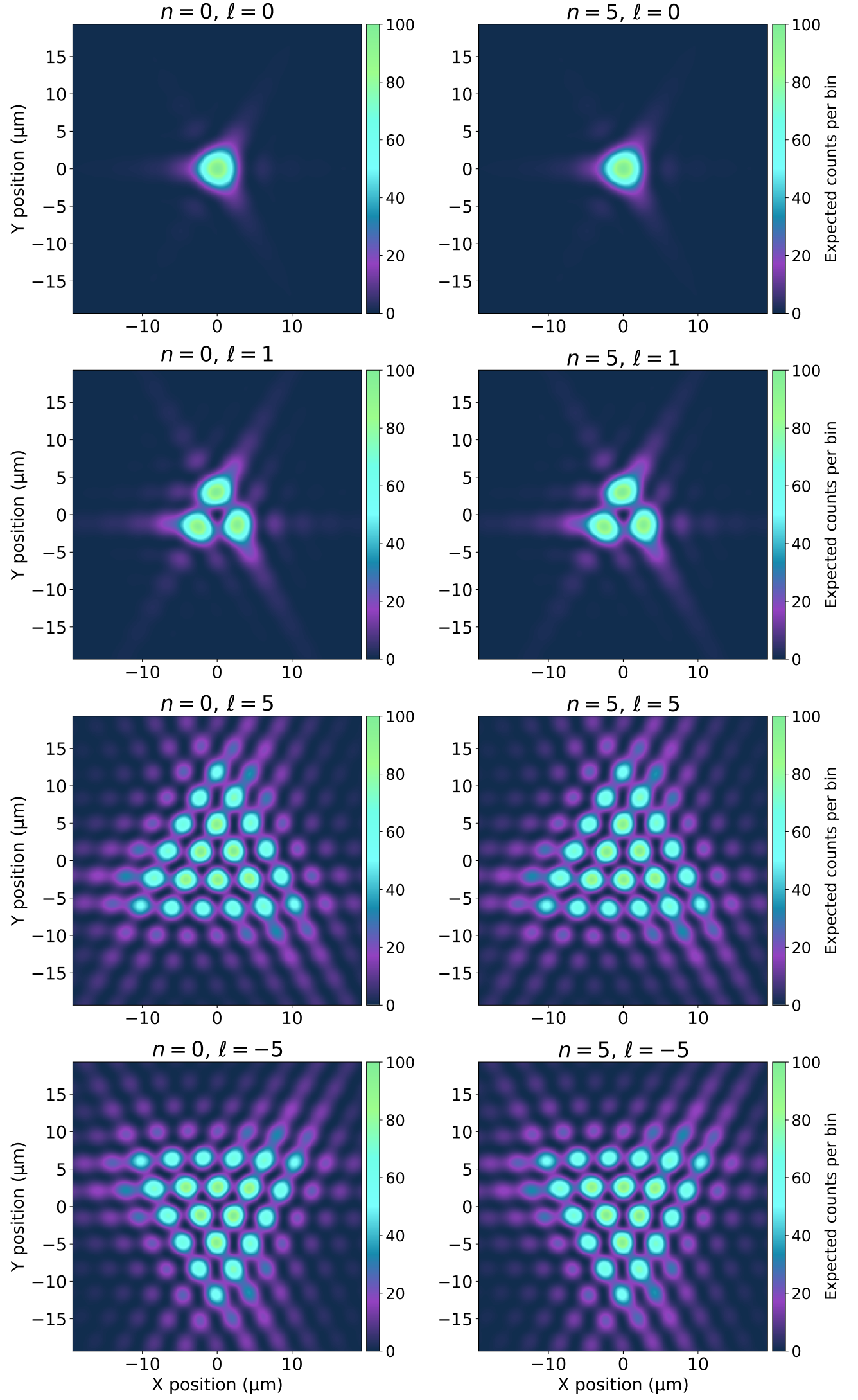}
\caption{Dependence on the radial index $n$ for LG packets through a triangular aperture (expected-count maps).
Same geometry as in Sec.~\ref{subsec:triangle-electrons}.
Left column: $n=0$; right column: $n=5$; rows: $\ell=0,1,5,-5$.
Varying $n$ changes the envelope/contrast; the OAM readout used in the main text is unchanged.}
\label{fig:LG-tri-nscan-app}
\end{figure}

Figure~\ref{fig:LG-circ-app} provides LG maps for the circular aperture used as a symmetry benchmark; the appearance is consistent with the Bessel case discussed in Sec.~\ref{subsec:circ}. Figure~\ref{fig:LG-tri-nscan-app} shows an $n$-scan for the triangular aperture: varying $n$ chiefly affects the envelope/contrast, while the OAM readout features used in the main text remain unchanged. For completeness, we verified on additional panels (not shown) that changing the triangle side $L$ rescales the far-field pitch according to $\Delta\propto 1/L$ (cf.\ Sec.~\ref{sec:feasibility}) without altering the OAM-specific structure.

\section{Triangular aperture: full derivations}\label{app:triangle-derivations}
We use screen coordinates $(x,y,z)$ and aperture coordinates $(x',y',0)$ throughout.
Equations in the main text (Secs. ~\ref{subsec:fraunhofer-mapping}, ~\ref{subsec:triangle-analysis}) already provide the Kirchhoff–Fresnel form,
the Fraunhofer mapping, and the $(x,y)\!\leftrightarrow\!(k_x,k_y)$ identification; we do
not repeat them here.

\subsection*{Vortex input as a transverse differential operator}

For a vortex of charge $\ell$ with a slowly varying envelope $u(x',y')$,
\[
\psi_{\rm in}(x',y') = \rho'^{\,|\ell|} e^{i\ell\varphi'}\,u(x',y')
=
\begin{cases}
(x'+iy')^{\ell}\,u, &\ell\ge 0,\\[2pt]
(x'-iy')^{|\ell|}\,u, &\ell<0,
\end{cases}
\]
and the identity
$\;\mathcal F\!\big[x'^m y'^n A(x',y')\big]
= i^{\,m+n}\partial_{k_x}^{\,m}\partial_{k_y}^{\,n}\,\tilde A(\mathbf k)\;$ gives
\begin{equation}
\widetilde{A\,\psi_{\rm in}}(\mathbf k)
\;\propto\;
\sum_{m=0}^{|\ell|}
\binom{|\ell|}{m}(\pm i)^{\,|\ell|-m}\,i^{\,m}\,
\partial_{k_x}^{\,m}\partial_{k_y}^{\,|\ell|-m}\,\tilde A(\mathbf k),
\label{eq:diff-op}
\end{equation}
with the upper (lower) sign for $\ell\!\ge\!0$ ($\ell\!<\!0$). Thus the vortex acts as an
$|\ell|$-th order differential operator in $\mathbf k$-space, enhancing rapidly varying
regions of $\tilde A$ as bright spots. 

The number of such spots equals the number of nonnegative pairs $(m,n)$ with
$m+n\le|\ell|$:
\begin{equation}
N(|\ell|)=
\sum_{m=0}^{|\ell|}\;\sum_{n=0}^{|\ell|-m} 1
= \frac{\bigl(|\ell|+1\bigr)\bigl(|\ell|+2\bigr)}{2}.
\label{eq:N-ell}
\end{equation}
This counting is aperture-independent in generic cases (barring accidental cancellations); the \textit{positions} are set by the mask.

\subsection*{Fourier transform of a triangular mask}

Consider a filled triangle with vertices \(\mathbf v_0,\mathbf v_1,\mathbf v_2\) in
the aperture plane \(z=0\). Introduce the edge vectors
\begin{equation}
\mathbf e_1 = \mathbf v_1 - \mathbf v_0, 
\qquad 
\mathbf e_2 = \mathbf v_2 - \mathbf v_0.
\end{equation}
The Fourier amplitude of the mask is
\begin{equation}
\label{eq:TriFT-def_app}
\tilde A(\mathbf k) = \iint_{\triangle} e^{-i\mathbf k\cdot\mathbf r}\,d^2\mathbf r , 
\qquad \mathbf k=(k_x,k_y).
\end{equation}

A convenient parameterization of the interior of the triangle is given by the
2-simplex mapping
\begin{equation}
\mathbf r(s,t) = \mathbf v_0 + s\,\mathbf e_1 + t\,\mathbf e_2,
\, s\ge 0,\; t\ge 0,\; s+t\le 1,
\end{equation}
with Jacobian
\begin{equation}
J = |\mathbf e_1\times \mathbf e_2|,
\end{equation}
which equals twice the area of the triangle.
Introducing the linear forms
\begin{equation}
\alpha = \mathbf k\!\cdot\!\mathbf e_1, 
\qquad 
\beta = \mathbf k\!\cdot\!\mathbf e_2,
\end{equation}
the integral becomes
\begin{equation}
\label{eq:TriFT-start_app}
\begin{aligned}
\tilde A(\mathbf k)
&= J\, e^{-i\mathbf k\cdot \mathbf v_0}
\int_{0}^{1}\!ds\int_{0}^{1-s}\!dt\;e^{-i(\alpha s+\beta t)} .
\end{aligned}
\end{equation}

\begin{figure}[!h]
    \centering
    \includegraphics[width=\linewidth]{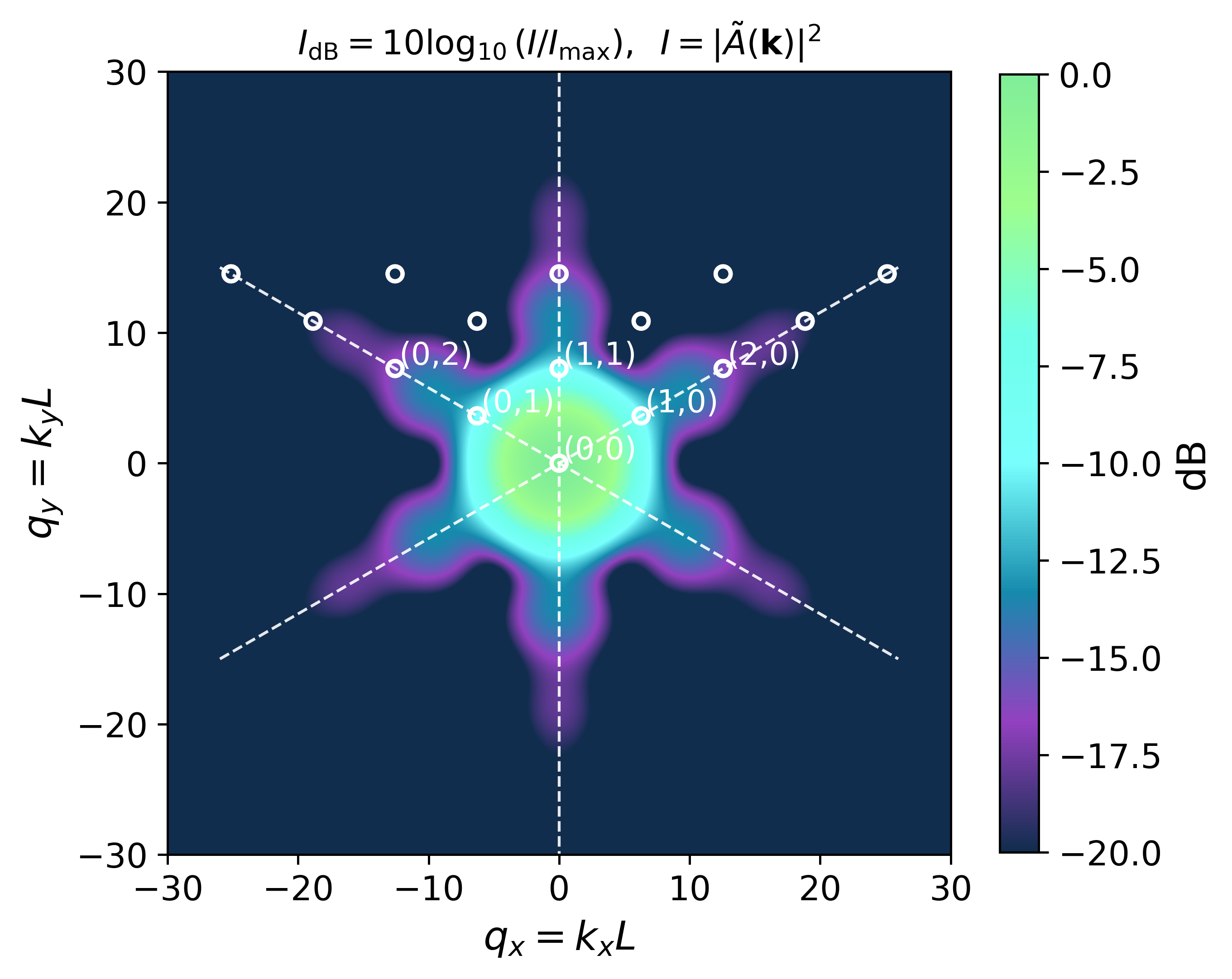}
    \caption{%
    Analytic Fourier amplitude of an equilateral triangular aperture (side $L=0.5~\mu$m).
    Color scale shows $S_{\rm dB}=10\log_{10}\!\big(|\tilde A(\mathbf k)|^2/\max_{\mathbf k}|\tilde A(\mathbf k)|^2\big)$; axes $q_x=k_xL$, $q_y=k_yL$ (dynamic range clipped at $-20$\,dB).
    Dashed lines indicate the three edge-orthogonal ridge families
    $\alpha=\mathbf k\!\cdot\!\mathbf e_1=0$, $\beta=\mathbf k\!\cdot\!\mathbf e_2=0$, and
    $\alpha-\beta=\mathbf k\!\cdot(\mathbf e_1-\mathbf e_2)=0$; their superposition yields the characteristic hexagonal symmetry.
    Open circles mark reciprocal-lattice nodes $\mathbf k_{m,n}=m\mathbf G_1+n\mathbf G_2$ selected by a vortex of charge $\ell$ with $m,n\ge0$ and $m+n\le|\ell|$ (here $|\ell|=4$; the sign of $\ell$ flips the orientation).
    Near the origin a few nodes are annotated by $(m,n)$.
    Note that $\tilde A(\mathbf k)$ itself corresponds to plane-wave illumination ($\ell=0$); the $\ell\neq0$ emphasis of these nodes arises from the $|\ell|$-th order derivatives acting on $\tilde A(\mathbf k)$ discussed in the text.%
    }
    \label{fig:triangle_fourie}
\end{figure}

Carrying out the $t$– and $s$–integrals gives a closed rational form (Eq.~\eqref{eq:triangle-rational}) quoted in the main text. This expression is exact and (up to relabeling of $\mathbf e_1,\mathbf e_2,\mathbf e_1-\mathbf e_2$) symmetric under permutation of the three edges. The linear forms
$\alpha$, $\beta$, and $\alpha-\beta=\mathbf k\!\cdot(\mathbf e_1-\mathbf e_2)$
pick out three preferred directions in Fourier space, each orthogonal to an
edge. Along these directions the pattern develops pronounced ridge lines of
alternating maxima and minima. Their superposition generates the characteristic
sixfold (hexagonal) symmetry of the far-field intensity (Fig.~\ref{fig:triangle_fourie}).

\subsection*{Reciprocal basis and node selection}

Define $\mathbf e_i\!\cdot\!\mathbf G_j=2\pi\delta_{ij}$ $(i,j=1,2)$. For an equilateral triangle of side $L$,
$|\mathbf G_1|=|\mathbf G_2|=4\pi/(\sqrt{3}\,L)$ and $\angle(\mathbf G_1,\mathbf G_2)=60^\circ$.
The three constructive-phase families
\[
\alpha=2\pi m,\qquad \beta=2\pi n,\qquad \alpha-\beta=2\pi p,
\qquad m,n,p\in\mathbb Z,
\]
intersect at the reciprocal-lattice nodes
\begin{equation}
\mathbf k_{m,n}=m\,\mathbf G_1+n\,\mathbf G_2,\qquad m,n\in\mathbb Z,
\label{eq:kmn}
\end{equation}
which map to detector coordinates via the Fraunhofer rule in Eq.~\eqref{eq:XY-from-k}.
The nearest-neighbour spacing on the screen is the lattice pitch
$\Delta$ quoted in Eq.~\eqref{eq:Delta-step} (see also Sec.~\ref{sec:feasibility}).

Acting with the $|\ell|$-th order operator \eqref{eq:diff-op} emphasizes a finite subset of
nodes; in particular, for generic orientations the nodes with $m,n\ge 0$ and $m+n\le|\ell|$
are highlighted. The global orientation flips with $\mathrm{sign}(\ell)$.

\subsection*{Optimization at fixed detector pitch}
\label{app:opt-derivation}

We derive the optimum quoted in the main text from the Fraunhofer mapping and the scaling of the triangular mask, using the definitions introduced in App.~\ref{app:triangle-derivations} and Sec.~\ref{sec:theory}, ~\ref{sec:feasibility}.

\paragraph*{Fourier scaling and node values.}
From the definition of the mask transform [Eq.~\eqref{eq:TriFT-def_app}] and the change of variables $\mathbf r=L\mathbf u$,
\begin{equation}
\tilde A_L(\mathbf k)=L^{2}\,\tilde A_1(L\mathbf k).
\label{eq:app-scaling}
\end{equation}
For the triangular reciprocal lattice [Eq.~\eqref{eq:kmn}], one has
$|\mathbf G_{1}|=|\mathbf G_{2}|=4\pi/(\sqrt{3}\,L)$, hence
$L\,|\mathbf G_{m,n}|$ is $L$–independent. Evaluated at a node,
\begin{equation}
\big|\tilde A_L(\mathbf G_{m,n})\big| = C_{\rm shape}\,L^{2},
\label{eq:app-A-node}
\end{equation}
where $C_{\rm shape}=O(1)$ depends on $(m,n)$ and the triangle geometry (but not on $L$).

\paragraph*{Detector-plane intensity at nodes.}
In the Fraunhofer zone the scalar field on the detector takes the standard far-field form
\begin{equation}
\begin{aligned}
\psi(x,y,z)
&= \frac{e^{ik\left(z+\tfrac{x^2+y^2}{2z}\right)}}{i\,\lambda_{\rm dB}\,z}\,
\iint_{S} A(x',y')\,\psi_{\rm in}(x',y')\\
&\hspace{3.6em}\times\,
\exp\!\left[-\,i\,\frac{k}{z}\,\big(x\,x' + y\,y'\big)\right]\;dx'\,dy'.
\end{aligned}
\label{eq:app-FF-field}
\end{equation}
Using the transform definition [Eq.~\eqref{eq:TriFT-def_app}] together with the detector–Fourier mapping of Eq.~\eqref{eq:kxky-mapping}, this becomes
\begin{equation}
\psi(x,y,z)
= \frac{e^{ik\left(z+\tfrac{x^2+y^2}{2z}\right)}}{i\,\lambda_{\rm dB}\,z}\;
\widetilde{A\psi_{\rm in}}(k_x,k_y).
\label{eq:app-FF-FT}
\end{equation}
Therefore the intensity is
\begin{equation}
I(x,y,z)\;\equiv\;|\psi(x,y,z)|^{2}
=\frac{1}{\lambda_{\rm dB}^{2}\,z^{2}}\,
\big|\widetilde{A\psi_{\rm in}}(k_x,k_y)\big|^{2}.
\label{eq:app-I}
\end{equation}
Near a reciprocal-lattice node, for a hard mask with a slowly varying incident envelope, we use the local factorization
\[
\widetilde{A\psi_{\rm in}}(k_x,k_y)\approx \tilde A(k_x,k_y)\,F_{\rm in},
\]
where \(F_{\rm in}=O(1)\) and is locally independent of \(L\). Evaluating \eqref{eq:app-I} at a node and using the node scaling from Eq.~\eqref{eq:app-A-node} gives
\begin{equation}
I_{\rm node}(L,z)=\frac{L^{4}}{\lambda_{\rm dB}^{2}\,z^{2}}\,|F_{\rm in}|^{2},
\label{eq:app-Inode}
\end{equation}
and eliminating \(z\) via Eq.~\eqref{eq:Delta-step} yields
\begin{equation}
I_{\rm node}(\Delta,L)=\frac{L^{2}}{\Delta^{2}}\,|F_{\rm in}|^{2},
\label{eq:app-Inode-Delta}
\end{equation}
i.e., at fixed \(\Delta\) the node brightness increases monotonically with \(L\).

\paragraph*{Fraunhofer bound and optimum.}
Imposing the far-field condition [Eq.~\eqref{eq:fraunhofer}] and substituting $z=(\sqrt{3}/2)(\Delta L/\lambda_{\rm dB})$ from Eq.~\eqref{eq:Delta-step} yields the upper bound
\begin{equation}
L \le \frac{\sqrt{3}}{2\,C}\,\Delta.
\label{eq:app-L-bound}
\end{equation}
Since \eqref{eq:app-Inode-Delta} is monotone in $L$, the maximum occurs at the boundary \eqref{eq:app-L-bound}, which reproduces the optimum stated in the main text (Eq.~\eqref{eq:opt-Lz}) and the corresponding optimized intensity $I_{\rm opt}\propto 3/(4C^{2})\,|F_{\rm in}|^{2}$ quoted there.


\bibliographystyle{apsrev4-2}
\bibliography{Citations_Diffraction.bib}

\end{document}